\documentclass[sigconf, screen]{acmart} %anonymous for double blind review to hide the authors infos

\usepackage{graphicx}
\usepackage{listings}
\usepackage{tabularx}
\usepackage{booktabs}
\usepackage{multirow}
\usepackage{makecell}
\usepackage{tikz}
\usepackage{hyperref}
\usepackage{verbatim}

\usepackage{xcolor}
\usepackage{tcolorbox}
\tcbuselibrary{breakable, skins}

\usepackage{soul}

\newlength{\vspacesarebad}
\def\structskip{\vskip 3pt plus 1pt minus 3pt} % 3+1-1 = smallskip
\def\invisibleskip{\vskip 1pt plus 0pt minus 1pt}

\makeatletter
\newcommand{\reviewonly}[1]{%
  \if@ACM@anonymous
    #1%
  \fi
}
\makeatother

\tcbset{boxsep=-1pt,left=2ex,right=2ex,before=\structskip\par\noindent,breakable, enhanced}

\lstdefinelanguage{yaml}{
  sensitive=false,
  comment=[l]{\#},
  morestring=[b]",
  morestring=[b]',
  morekeywords={
    patterns,name,body,engine,language, id, rule, pattern, fix,where
  },
  keywordstyle=\color{blue}\bfseries
}

\lstdefinelanguage{toml}{
  keywords={true,false},
  morekeywords={[2]rule},
  keywordstyle=[2]\color{blue}\bfseries,
  sensitive=true,
  comment=[l]{\#},
  morestring=[b]",
  morestring=[b]',
  alsoletter={-}
}

\newcommand{\blue}[1]{{\color{blue}{#1}}}
\newcommand{\orange}[1]{{\color{blue}{#1}}}

\newif\ifCameraReady
\newif\ifBlueHighlights
\CameraReadytrue %% <---- a commenter/decommenter ici seulement !!!
\newcommand\clarification[1]{{#1}}
\newcommand\removed[1]{\textcolor{red}{\st{#1}}}
\newcommand\camera[1]{{#1}}

\ifCameraReady
\renewcommand\removed[1]{}
\fi
\ifBlueHighlights
\renewcommand\clarification[1]{\textcolor{blue}{#1}}
\renewcommand\camera[1]{\textcolor{blue}{#1}}
\else

\renewcommand{\blue}[1]{{#1}}
\renewcommand{\orange}[1]{{#1}}
\fi

\setcopyright{cc}
\setcctype{by}
\acmDOI{10.1145/3832783.3837504}
\acmYear{2026}
\copyrightyear{2026}
\acmISBN{979-8-4007-2882-2/2026/10}
\acmConference[ASE '26]{Proceedings of the 41st IEEE/ACM International Conference on Automated Software Engineering}{October 12--16, 2026}{Munich, Germany}
\acmBooktitle{Proceedings of the 41st IEEE/ACM International Conference on Automated Software Engineering (ASE '26), October 12--16, 2026, Munich, Germany}
\acmSubmissionID{ase26main-p1631-p}
\received{2026-03-26}
\received[accepted]{2026-06-18}

\title{Code Transformation Rule Synthesis using LLMs: Potential and Limits}

\author{Axel Allain}
\correspondingauthor
\orcid{0009-0009-9088-6178}
\affiliation{
  \institution{Univ Rennes, INRIA, CNRS, IRISA}
  \city{Rennes}
  \country{France}
}
\email{axel.allain@inria.fr}

\author{Aymeric Blot}
\orcid{0000-0003-0485-5279}
\affiliation{
  \institution{Univ Rennes, INRIA, CNRS, IRISA}
  \city{Rennes}
  \country{France}
}
\email{aymeric.blot@irisa.fr}

\author{Djamel Eddine Khelladi}
\orcid{0000-0002-2218-650X}
\affiliation{
  \institution{Univ Rennes, INRIA, CNRS, IRISA}
  \city{Rennes}
  \country{France}
}
\email{djamel-eddine.khelladi@irisa.fr}

\author{Mathieu Acher}
\orcid{0000-0003-1483-3858}
\affiliation{
  \institution{Univ Rennes, INRIA, CNRS, IRISA}
  \city{Rennes}
  \country{France}
}
\email{mathieu.acher@irisa.fr}

\date{February 2026}

\begin{document}

\begin{abstract}

%%Large Language Models (LLMs)
%LLMs are becoming increasingly effective in software evolution tasks such as automated program repair, refactoring, API misuse fixing, and code migration. 
%These advances can substantially reduce the manual effort required for maintaining, refactoring, and evolving complex software systems. 
%LLMs are becoming increasingly effective in code generation and transformation.  Still, they 

Due to their black-box nature, LLMs suffer from limited explainability and a lack of determinism. 
Their usage cost can also rise, particularly with repetitive tasks on large codebases. 
%Code transformation rules mitigate those challenges \reviewonly{ by providing a high degree of explainability, determinism, and reuse}. % on large codebase. 
%Hence, an opportunity is to leverage LLMs to generate the code transformation rules rather than directly updating the code. 
%This paper fills this gap, with a novel empirical study targeting
To mitigate this, we conduct a novel empirical study targeting
%Instead of focusing solely on direct code generation, we investigate whether LLMs can produce explicit and reusable transformation rules expressed 
three domain-specific languages for transformation rules, namely Comby, GritQL, and Ast-Grep.
We evaluate three LLMs (GPT-5.4, GPT-oss-120B, and Llama3.1-8B) on six diverse datasets covering four software-evolution tasks: API misuse correction, program repair, API migration, and language version migration. %We analyze how such rules can be derived from code changes and 
\reviewonly{We evaluate the syntactic and semantic validity, structural completeness, generalization capacity, and transformation correctness for LLM-generated rules. We complement this quantitative evaluation with a qualitative analysis of sampled success and failure cases.}
Our results provide evidence that transformation rule synthesis moves beyond proof-of-concept with strong frontier models. 
GPT-5.4 achieves consistently high rule applicability rates and produces transformations closest to the ground truth across most benchmarks. Smaller and open-weight GPT-oss-120B and Llama3.1-8B models remain effective for simpler, localized changes but struggle with complex migration scenarios. \blue{We also observe non-negligible generalizability through the usage of meta-variables and through a high reuse score in the first quartile of many datasets.
%However, GPT-5.4 effectiveness depends strongly on the software-evolution task: localized edits such as API misuse corrections are substantially more amenable to general, reusable rule synthesis than function-level repairs or multi-statement migrations.
Finally, when compared to the anti-unification algorithm, LLMs outperform it in correctness, but underperform  in rule applicability. %on most evaluation metrics. 
Overall, our results show great potential for LLMs to generate sound, correct, generalizable, and reusable rules.
} 
%Overall, our quantitative and qualitative analyses show that the main question has shifted from whether LLMs can synthesize transformation rules to under which task and conditions the generated rules are sound, correct, generalizable, and reusable.
\end{abstract}




\maketitle

\section{Introduction}
Software continuously evolves in scale and complexity to accommodate technological advances and emerging use cases~\cite{siy1998challenges,sarkar2009software,northrop2006ultra}. 
Many challenges arise from the costly phase of software maintenance and evolution~\cite{glass2001frequently,alkhatib1992maintenance,banker1993software}.
For example, as APIs and programming frameworks evolve, developers must continuously evolve their code and repeatedly cope with various tasks, such as program repair, refactoring, API misuse fixing, code migration, co-evolution, etc. 
These evolution tasks are tedious, error-prone, time-consuming, and drive up the costs of maintenance. Therefore, automating these laborious tasks is highly beneficial for developers, as it increases efficiency and allows them to focus on higher-value development efforts. 
Rich literature exists on automating the above tasks~\cite{le2019automated,liu2021critical,monperrus2018living,golubev2021one,lacerda2020code,mens2004survey,amann2016mubench,amann2018systematic,sven2019investigating,li2021large,nguyen2016mapping,khelladi2020co,le2021untangling,kebaili2025automated,miranda2025test}, and
recent advances show that LLMs are becoming increasingly effective in resolving them~\cite{cordeiro2024empirical,ziftci2025migrating,kebaili2024empirical,zine2025llm,zhang2023multilingual,bouzenia2025repairagent,jin2023inferfix,yang2025survey,cordeiro2025llm,shirafuji2023refactoring}. 

However, as the code is generated directly by the LLMs, it still faces major challenges. Indeed, LLMs lack explainability,  determinism, and systematic reuse with reduced costs. The black-box behavior of LLMs limits their adoption in critical systems and large-scale environments as they need safe code transformations that are explainable and deterministic. Moreover, in large-scale codebase %refactoring scenarios, 
the use of LLMs for systematic code transformation %across %extensive codebase
can lead to increased costs, particularly for organizations with limited resources.   

To mitigate these challenges, developers can rely on code transformation DSLs to write transformation rules, which provide a high degree of explainability, determinism, and reuse on large codebase. 
Yet, transformation rules written in existing domain specific languages (DSLs) such as  Comby, GritQl, and Ast-Grep often have a steep learning curve and require expertise to use effectively. Writing and maintaining these rules can be tedious, time-consuming, and error-prone because of their rigid nature, limiting their usability across projects.
Hence, an opportunity arises to leverage LLMs to generate the code transformation rules rather than directly updating the code. Yet, to the best of our knowledge, there is a lack of work on LLMs' ability to formalize and generate code transformation rules. We fill this gap.

In this paper, we present a novel empirical study on the ability of LLMs to generate transformation rules by inferring DSL-based rewrite patterns from observed code changes, i.e., pairs of original and evolved code.
\removed{In particular, on three state-of-the-art DSLs, namely Comby, GritQL, and Ast-Grep.}
Due to large-scale pretraining on code transformation corpora, LLMs have developed a strong ability to learn semantic representations of code changes, including API mappings, program repair strategies, refactoring,  and migration/co-evolution patterns. To support rule generation, we provide the LLM with representative examples of documentation and transformation rules from different DSLs, allowing us to assess how additional context influences its ability to infer correct transformation rules.
\removed{We rely on prompt engineering to guide LLMs in generating transformation rules at the file level.}
We also experiment with augmenting prompts using examples of documentation and rules, including a limited use of retrieval for one DSL (Comby), to assess its impact on rule generation.

We evaluate the capability of three LLMs (\textit{GPT-5.4}, \textit{GPT-oss-120B}, and \textit{Llama3.1-8B}) to generate transformation rules across four software engineering tasks, program repair, API misuse correction, API migration, and language version migration, on six dedicated benchmarks. Our goal is to identify which DSLs and types of transformation rules are most effectively handled by LLMs and which remain the most challenging. Our results provide evidence that rule synthesis moves beyond proof-of-concept with strong frontier models (\textit{GPT-5.4}), while revealing that effectiveness depends strongly on the software-evolution task and the fit between the edit structure and the rule abstraction. \camera{We also found that the generated rules can abstract code transformations through the use of metavariables, resulting in high reuse scores across many datasets. Moreover, compared with the anti-unification baseline, LLMs produce more correct and semantically meaningful rules while avoiding the over-generalization inherent to the anti-unification algorithm. Finally, test-based validation on the program repair benchmarks showed that only a small fraction of the generated transformations that failed AST matching successfully repaired bugs.}

\structskip
In summary, our novel contributions are as follows:
\newline\textbf{A systematic evaluation} of three LLMs on transformation rule synthesis, analyzing their soundness, correctness, and generalization across three DSLs, four code evolution tasks, and six datasets.
\newline\textbf{A comparative analysis} across three transformation rule DSLs, providing evidence on the extent to which LLMs understand these languages, including the effect of retrieval-augmented generation (RAG) on rule quality.
\newline\textbf{A qualitative analysis} of success and failure modes that explains the concrete mechanisms behind model differences, including formatting failures, over- and under-generalization, and missing edits.
\newline\textbf{Insights on task suitability for rule synthesis}: showing that localized edits are substantially more amenable to general, reusable rule synthesis than function-level or multi-statement edits.
\newline\textbf{A replication package} is available at \orange{{\small{\url{https://anonymous.4open.science/r/EmpiricalStudyTransformationRules-13B3}}}}

\section{Background}
\removed{To avoid developers burden in editing recurrent changes in large codebase, code transformation tools/DSLs allow the definition of pre-written code transformation rules that can be applied recursively. These transformations techniques are popular to automate code evolution tasks such as API migration, API misuse, refactoring, language version migration and program repair.
However, the usage of grep, sed, sd or ripgrep with regular expressions to locate and modify code fragments requires good knowledge with a steep learning curve. Also, these techniques operate purely at the textual level and sometimes fail to capture the syntactic structure of programs, which may lead to incorrect transformations.
To overcome these limitations,  developed DSLs made transformation rules easier to write using metavariables in matching patterns. 
Indeed, most DSLs are made of two main parts for \texttt{matching} and \texttt{rewriting} code. The matching pattern will either match the code syntax or the code AST. For that, it will use placeholders acting as metavariables that capture program elements like AST nodes or pure syntax fragments. The rewriting pattern can reuse those metavariables defined in the matching pattern to transform the matched code, keeping only relevant parts of the modification and simplifying the rule structure. Several tools implement rule-based transformation DSLs, such as Coccinelle for C programs targeting Linux co-evolution tasks, Comby for structural code search and rewrite, and AST-based tools like Ast-Grep and GritQL.}
\orange{To reduce the burden of recurrent changes in large codebase, developers often rely on code search and transformation tools. Search tools such as \texttt{grep} and \texttt{ripgrep} efficiently locate code fragments, while transformation tools such as \texttt{sed} and \texttt{sd} automate repetitive code edits. These tools help developers perform large-scale modifications while reducing manual effort.}
\orange{However, using tools such as \texttt{sed}, \texttt{sd}, or regular expressions requires substantial expertise and often involves a steep learning curve. Moreover, these techniques operate purely at the textual level and cannot capture the syntactic structure of programs, which may lead to incorrect or unintended transformations.}
\orange{To overcome these limitations, dedicated code transformation DSLs have been developed to make transformation rules easier to write and reuse. These DSLs typically rely on matching patterns enriched with metavariables, enabling developers to express transformations at the syntactic or AST level rather than as plain text. Most DSLs consist of two main components: a \texttt{matching} pattern and a \texttt{rewriting} pattern. The matching pattern captures program elements, such as syntax fragments or AST nodes, using metavariables. The rewriting pattern then reuses these metavariables to generate the transformed code, preserving relevant program elements while simplifying the specification of transformation rules. Several tools implement such rule-based transformation DSLs, including Coccinelle for C programs and Linux kernel co-evolution, Comby for structural code search and rewrite, and AST-based tools such as Ast-Grep and GritQL.}

\section{Motivating Example}

Large-scale codebase often require developers to apply repetitive transformations across many files. Manually performing these modifications is tedious and error-prone, while repeatedly prompting large language models (LLMs) to perform each individual change may increase computational and token costs. At the same time, LLMs remain vulnerable to hallucination issues which may introduce errors during the code modification process. Hence, leading to a lack of explainability and determinism in code evolution.  

However, one can address these challenges with code transformation rules. 
\removed{For example, consider the refactoring where a deprecated logging API is replaced with a new structured logging interface. The following illustrates simple instances of such a transformation.}
\orange{As an illustrative example, consider the diff below.}

\newcommand{\lstbg}[3][0pt]{{\fboxsep#1\colorbox{#2}{\strut #3}}}
\lstdefinelanguage{diff}{
  morecomment=[f][\lstbg{red!20}]-,
  morecomment=[f][\lstbg{green!20}]+,
}

\ifCameraReady\else
{\small
\begin{lstlisting}[
language=diff,
basicstyle=\ttfamily\footnotesize\color{red},
keywordstyle=\color{red},
commentstyle=\color{red},
stringstyle=\color{red},
identifierstyle=\color{red}
]
- logger.log("User " + id + " logged in");
- logger.log("User " + user.name + " logged in");
- logger.log("User " + .... + " logged in");
+ logger.info("User {} logged in", id);
+ logger.info("User {} logged in", user.name);
+ logger.info("User {} logged in", ....);
\end{lstlisting}
}
\fi

{\small
\ifBlueHighlights\color{blue}\fi
\begin{lstlisting}[language=diff]
- if (user != null) { sendEmail(user); }
+ Optional.ofNullable(user)
+     .ifPresent(this::sendEmail);
    
\end{lstlisting}
}

\removed{This change modifies the logging method and replaces string concatenation with parameterized logging. While simple, this transformation may appear hundreds or thousands of times across a large project.}

\orange{This change replaces an explicit null checks with the more concise Optional API. While the transformation is simple, it may need to be applied in various contexts hundreds or thousands of times across a large project, making it well suited for automated refactoring.}
Using a rule-based transformation DSL such as Comby, this refactoring can be expressed as a reusable transformation rule:

\ifCameraReady\else
{
\begin{lstlisting}[
language=toml,
basicstyle=\ttfamily\footnotesize\color{red},
keywordstyle=\color{red},
commentstyle=\color{red},
stringstyle=\color{red},
identifierstyle=\color{red}
]
[newLoggingMethod]
match = logger.log("User " + :[var] + " logged in")
rewrite = logger.info("User {} logged in", :[var])
\end{lstlisting}
}
\fi

{
\ifBlueHighlights\color{blue}\fi
\begin{lstlisting}[language=toml]
[optionalIfPresent]
match = "if (:[var] != null) { :[func](:[var]); }"
rewrite = "Optional.ofNullable(:[var])
                    .ifPresent(this:::[func]);"
\end{lstlisting}
}

Once defined, the rule can be applied automatically across the entire codebase, transforming all matching occurrences in a single pass. 
The advantage of rule-based code transformations becomes more apparent at scale. If a change occurs $n$ times in a project, an LLM-based workflow may require either repeatedly prompting the model for each occurrence or providing large contexts containing many code fragments. Both approaches increase token consumption and latency, driving up the costs. In contrast, a transformation rule is written once and applied automatically to all matches, providing a deterministic, explainable, and efficient solution.

Recent advances in LLMs like GPT or Claude Code suggest that they may assist developers in generating such transformation rules automatically. Our hypothesis is that given pairs of code snippets representing before and after versions of a change, an LLM could potentially synthesize a generalized rule that captures the transformation pattern. 
However, it remains unclear whether LLMs can reliably synthesize correct and reusable transformation rules from example changes. Furthermore, many transformation DSLs exist, each with different syntactic and structural matching capabilities, and to the best of our knowledge, there has been no systematic evaluation of how well LLMs can generate transformation rules for them.
In this paper, we address this gap through a novel empirical study that evaluates the ability of LLMs to synthesize transformation rules from example code changes across three popular rule-based transformation tools: Comby, Ast-Grep, and GritQL.

\section{Research Questions}
With this empirical study we aim to answer the following four RQs:

\structskip % à fine tuner dans main.tex
\noindent \textbf{RQ1: To what extent can LLMs generate sound transformation rules from code diffs while maintaining reasonable computational and financial cost?} This assesses the ability of different LLMs to infer valid transformation rules, considering the costs of generating rules in terms of token expense.

\invisibleskip % à fine tuner dans main.tex
\noindent \textbf{RQ2: How complete are the generated transformation rules in terms of coverage, structural richness, and generalization capacity?} This aims to cover whether the generated rules capture the code changes in the datasets generalizing beyond individual examples using metavariables.

\invisibleskip % à fine tuner dans main.tex
\noindent \textbf{RQ3: How well do the generated transformation rules reproduce the ground-truth transformations in terms of semantic and syntactic similarity?} This evaluates how closely the transformations produced by the generated rules match the expected code modifications, considering the structural similarity of the intended program semantics.

\invisibleskip % à fine tuner dans main.tex
\noindent \textbf{RQ4: To what extent do different LLMs generate complementary transformation rules, and how much overlap exists between them?} This evaluates the overlap and uniqueness of rules across models to understand whether they capture similar or distinct transformation patterns.

\section{Methodology}

\begin{figure}[t]
\centering
\includegraphics[width=\linewidth]{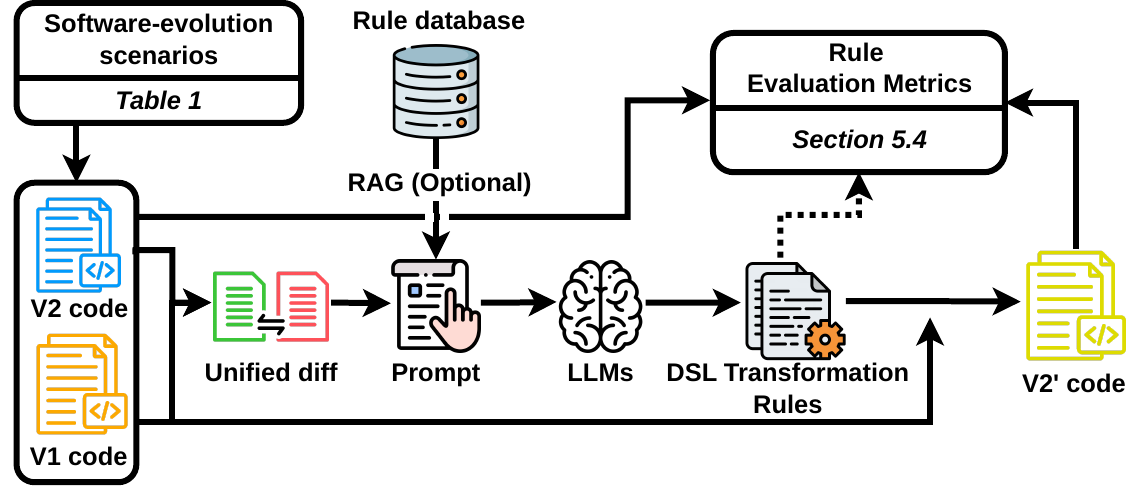}
\caption{High-level overview of the pipeline for rule generation and application to V1 file.}
\Description{High-level overview of the pipeline for rule generation and application to a V1 file.}
\label{fig:flow}
\end{figure}

To answer these questions, we propose the overall approach shown in \autoref{fig:flow}. It considers different software evolution scenarios. Based on code diffs and examples, LLMs generate transformation rules to replay the code evolutions, which are then evaluated with several quality metrics. \camera{Throughout the study, we followed when relevant the proposed guidelines for conducting empirical software engineering studies involving LLMs from Baltes et al. \citep{baltes_guidelines_2026}.}

\subsection{Models}

We consider three large language models with different architectures and capabilities: \textit{GPT-5.4}, \textit{GPT-oss-120B}, and \textit{Llama3.1-8B}. GPT-5.4 is a state-of-the-art proprietary model known for its strong reasoning ability and large context window, making it well-suited for complex code transformation and rule inference tasks. GPT-oss-120B is a recent open-weight model with 120 billion parameters and a context window of 131,072 tokens, designed to provide strong reasoning performance while remaining accessible for research experimentation. Finally, Llama3.1-8B is a smaller open-weight model with 8 billion parameters, included to provide a more lightweight and widely deployable model alternative.
\orange{We selected these models to cover diverse ecosystems, parameter scales, and deployment constraints, while reflecting strong coding and instruction-following capabilities. They are widely used in SE/code-generation research.}

\removed{For all three models, we set the temperature hyperparameter to 0 in order to obtain deterministic outputs and improve experiment reproducibility, which is important for consistently generating transformation rules.}

\orange{For all three models, we set the temperature to 0 to reduce output variability and improve experiment reproducibility. Although this does not guarantee fully deterministic outputs across executions, it generally leads to more consistent transformation rules.}

\subsection{DSL Tools of Code Transformers}

We selected three state-of-the-art DSLs for code transformations. 

\structskip % à fine tuner dans main.tex
\noindent
\textbf{Comby}: Comby \cite{van_tonder_lightweight_2019} is a language-agnostic and lightweight DSL for matching or rewriting syntactic structures of a program’s parse tree using transformation rules. The key functionality of Comby rules is the use of template variables, which are holes in the match and rewrite templates that can be filled with code. Comby is also language-aware with an understanding of  basic syntax of code, strings, and comment syntax. 

For example, the Comby rule below replaces \texttt{fit\_transform} with \texttt{transform} on test data to avoid re-fitting. The metavariables \texttt{:[obj]} and \texttt{:[arg]} capture the receiver object and its argument, enabling the transformation to apply generically.

{\small
\begin{lstlisting}[language=toml]
[robust_scaler_test_transform]
match="scaled_x_test=:[obj].fit_transform(:[arg])"
rewrite="scaled_x_test=:[obj].transform(:[arg])"
\end{lstlisting}
}

\structskip % à fine tuner dans main.tex
\noindent
\textbf{Ast-Grep}: Ast-Grep \cite{astgrep2026} is a structural search-and-replace tool that operates on the program’s abstract syntax tree (AST). Its match and rewrite patterns rely on metavariables that correspond to AST nodes and sub-expressions. Unlike Comby, which performs syntax-aware matching over textual code blocks, Ast-Grep enforces structurally valid matches at the AST level. As a result, Ast-Grep provides stricter and more precise matching, as each metavariable must correspond to a well-formed AST node.

For example, the Ast-Grep rule below casts the input to \texttt{float32} before calling \texttt{fit\_transform}. The metavariables \texttt{\$LHS}, \texttt{\$OBJ}, and \texttt{\$ARG} capture the assignment target, the object, and the input argument, enabling a generic transformation.

\begin{lstlisting}[language=yaml]
id: cast-affinity-float32
language: python
rule:
  pattern: $LHS = $OBJ.fit_transform($ARG)
fix: $LHS = $OBJ.fit_transform($ARG.astype(np.float32))
\end{lstlisting}

\structskip % à fine tuner dans main.tex
\noindent
\textbf{GritQL}: GritQL \cite{gritql2026} is also a structural search-and-replace tool operating on AST using a unique syntax close to SQL. This DSL offers a precise and expressive way of writing patterns for safer code transformations.
Consequently, it can be harder to learn for a new user as it is a custom language with a stiff learning path and can be overkill for a simple find-replace task.

\removed{For example, the GritQL rule below detects incorrect usage of \texttt{predict} with an extra argument and rewrites it to the correct form. The metavariables \texttt{\$obj}, \texttt{\$X}, \texttt{\$y}, and \texttt{\$expected} capture the model, input data, unintended extra argument, and expected output, allowing the transformation to generalize across different contexts.}

\orange{For example, the GritQL rule below detects incorrect usage of a \texttt{predict} function. The metavariables \texttt{\$obj}, \texttt{\$X}, \texttt{\$y}, and \texttt{\$expected} capture the model, input data, unintended extra argument, and expected output, generalizing the transformation across contexts. The \texttt{where} clause further constrains the rule by requiring a metavariable to match a specific language construct before the rewrite is applied.}

{\ifBlueHighlights\color{blue}\fi
\begin{lstlisting}[language=yaml,breaklines=false]
patterns:
  - name: remove_second_predict_arg
    body: |
      engine marzano(0.1)
      language python
      `assert_array_equal($obj.predict($X, $y), 
      $expected)` =>
      `assert_array_equal($obj.predict($X), $expected)`
      where $y <: python`None`
\end{lstlisting}
}

\smallskip
While these three DSLs use different syntaxes, they share common foundational concepts,
%notably the notions of a match and a rewrite, as well as the use of metavariables to generalize transformation rules. 
their main differences lie in the level of abstraction at which they operate, i.e., either on concrete or abstract syntax. 
Comby relies on a template system %, consisting of match and rewrite patterns, where
with metavariables symbolized as \texttt{:[mv\_name]} capture strings. This provides strong language-agnosticism and ease of use. However, it comes with limited structural expressiveness. For example, it cannot distinguish between a variable and a function call when they share the same textual form, unless additional delimiters are present. 
In contrast, Ast-Grep and GritQL build on Tree-Sitter\footnote{https://tree-sitter.github.io/tree-sitter/} parsers to provide structural awareness, though they differ in how transformations are expressed. %Ast-Grep adopts a rule-based approach centered around \textit{rule} and \textit{fix} constructs. 
Its expressiveness is enhanced by support for hierarchical constraints such as inside, has, and follows, which allow users to match nodes based on their relationships  within the AST. 
GritQL, on the other hand, departs from rigid key–value structures and instead uses a more functional syntax, where the ''$=>$'' operator enables inline transformations. It also supports variable scoping, allowing information captured deep within nested structures to be reused at higher levels. 

%Its strength lies in modularity: features such as the bubble operator support advanced variable scoping, allowing information captured deep within nested structures to be reused at higher levels, such as in top-level definitions.

\subsection{Datasets}

This section details the six datasets used in our evaluation, grouped by type of software engineering task, for which we extracted human-written pairs of \emph{before} and \emph{after} code modification.
They are summarised in \autoref{tab:datasets}.
\clarification{We selected them according to six criteria: 1) state-of-the-art and open-access datasets, 2) diverse SE field origin, 3) availability of code pairs (input/output), 4) transformation granularity/complexity (small to large edits), 5) diversity of programming languages, 6) human-written edits (non-generated).}

\begin{table}[t]
\caption{%
Summary of the datasets used in the evaluations.}
\vspace{\vspacesarebad} % à modifier dans main.tex pour être consistant
\small
\begin{tabular}{l l r l}
\toprule
\textbf{Dataset} & \textbf{Language} & \textbf{Pairs} & \textbf{Typical Change Type} \\
\midrule
\multicolumn{4}{c}{\textbf{API Misuse}} \\
Galappaththi et al. & Python & 43 & Statement-level fixes \\
\cmidrule(lr){1-4}
\multicolumn{4}{c}{\textbf{Program Repair}} \\
ManySStuBs4J & Java & 528 & Single-statement bug fixes \\
Defects4J & Java & 427 & Function-level bug fixes \\
BugsInPy & Python & 559 & Function-level bug fixes \\
\cmidrule(lr){1-4}
\multicolumn{4}{c}{\textbf{API Migration}} \\
PyMigBench & Python & 605 & Multi-statement changes \\
\cmidrule(lr){1-4}
\multicolumn{4}{c}{\textbf{Language Version Migration}} \\
JMigBench & Java & 45 & Method-level changes \\
\bottomrule
\end{tabular}
\label{tab:datasets}
\vspace{\vspacesarebad} % à modifier dans main.tex pour être consistant
\end{table}

\structskip % à fine tuner dans main.tex
\textbf{(1) API Misuse}
%\newline 
\textit{Galappaththi et al.~\cite{galappaththi_empirical_2024}}: A dataset of 43 data-dependent API misuses collected from Stack Overflow posts and GitHub commits related to data science libraries.

\invisibleskip % à fine tuner dans main.tex
\textbf{(2) Program Repair}
%\newline 
\textit{ManySStuBs4J~\cite{karampatsis_how_2020}}: A corpus of simple statement-level Java bug fixes mined from open-source GitHub projects. It contains two variants: one from the top 100 Java Maven projects and one from the top 1000 Java projects. For this experiment, we selected a random subset of 528 pairs due to limited time and computational resources.
%\newline\indent 
\textit{Defects4J}~\cite{just_defects4j_2014}: A benchmark of 854 reproducible Java bugs from real-world open-source projects, including faulty and fixed versions along with test suites. For this experiment, we selected a random subset of 427 pairs due to limited time and computational resources.
%\newline\indent 
\textit{BugsInPy~\cite{widyasari_bugsinpy_2020}}: A benchmark of 559 real-world Python bugs from 17 projects, inspired by Defects4J, designed for controlled debugging and repair studies.

\invisibleskip % à fine tuner dans main.tex
\textbf{(3) API Migration}
%\newline 
\textit{PyMigBench~\cite{islam_pymigbench_2023}:} A benchmark of Python library migrations with 3,096 migration-related code changes from 335 migrations between 141 analogous library pairs. For our experiment, it represents 605 pairs of code.

\invisibleskip % à fine tuner dans main.tex
\textbf{(4) Language Version Migration}
%\newline 
\textit{JMigBench~\cite{amin_jmigbench_2026}:} A benchmark suite and evaluation pipeline designed to assess large language models (LLMs) in the task of migrating Java functions from Java~8 to Java~11. It contains 45 pairs of Java~8/Java~11 migrations.

\subsection{Evaluation Metrics}
\removed{To evaluate the effectiveness of the generated transformation rules, their characteristics, and the correctness of the resulting code transformations, we use the following metrics.}

%\structskip % à fine tuner dans main.tex
%\textbf{Number of Pairs (N)}: Number of (before, after) code pairs representing individual edits.

\orange{Soundness of the generated DSL transformation rules is assessed through metrics proposed by Ramos et al.~\cite{ramos_melt_2023} and Ketkar et al.~\cite{ketkar_inferring_2022}.}

%\invisibleskip % à fine tuner dans main.tex
\textbf{Failures (F)}: Number of code edits for which the generated transformation rule could not be successfully applied or evaluated. This includes cases where (i) the generated rule is syntactically invalid, (ii) the DSL engine fails during execution, (iii) the produced migrated code contains syntax errors, or (iv) metric computation (e.g., AST parsing) fails.

\invisibleskip % à fine tuner dans main.tex
\textbf{Rule Applicability Rate (RA\%)}: The Rule Applicability Rate is the percentage of generated transformation rules that are syntactically valid and executable by the corresponding DSL engine.

\invisibleskip % à fine tuner dans main.tex
\textbf{Number of Generated Tokens (NT)}: Measures the number of tokens generated by the LLM when producing a set of transformation rules for a given code edit.

\structskip % à fine tuner dans main.tex
\orange{Completeness of the generated DSL transformation rules is assessed using  metrics for characterizing efficiency and verbosity.}

%\invisibleskip % à fine tuner dans main.tex
\textbf{Number of Rules (\#R)}: Mean number of rules in a given DSL configuration file for one code edit.

\invisibleskip % à fine tuner dans main.tex
\textbf{Size of Rule (SR)}: Mean rule size for each dataset, as the length of the character sequence.

\invisibleskip % à fine tuner dans main.tex
\textbf{Metavariables Token Ratio (MTR)}: The MTR measures how much of the match pattern is hard-coded based on the number of metavariables tokens used in a single pattern.
\[
\mathrm{MTR} =
\frac{\#\,\text{metavariable tokens}}
{\#\,\text{total tokens in the pattern}},
\qquad
0 \le \mathrm{MTR} \le 1.
\]
    An MTR of 0 indicates that no metavariables are used (fully hard-coded pattern), 
    while an MTR of 1 indicates that every token in the pattern is a metavariable.

\structskip % à fine tuner dans main.tex
\orange{Finally, the following syntactic metrics are widely used for evaluating code similarity and transformation correctness.}

%\invisibleskip % à fine tuner dans main.tex
\textbf{Tree edit Distance (TD)}: The Tree edit Distance calculates the shortest sequence of edit operations (Insert, Delete, Replace) to transform tree \textit{T1} to tree \textit{T2}. In our case, \textit{T1} is the human-written reference code, while \textit{T2} corresponds to the code produced by applying the generated transformation rule.

\invisibleskip % à fine tuner dans main.tex
\textbf{Exact Match (EM\%)}: Exact Match is a binary metric for a syntax-level perfect match between the human-written reference code and the transformation rule-generated code.

\invisibleskip % à fine tuner dans main.tex
\textbf{AST Match (AM\%)}: AST Match is a binary metric for an AST-level perfect match between the human-written reference code and the transformation rule-generated code.

\subsection{Evaluation Process}
To evaluate each code change of the 6 datasets described above, we provide the LLM with a diff from a pair of code snippets representing the before and after versions of a modification, as shown in \autoref{fig:flow}. A diff allows a developer to look at the two files side by side and see exactly what differentiates them, such as new lines of code that have been added, if variable names have been changed, or if any lines of code have been removed. Using this input-output example, the model is prompted to synthesize one or more transformation rules in a chosen DSL: Comby, Ast-Grep, or GritQL. Since LLMs have limited exposure to these DSLs during training, we augment the prompt with 8 examples written in the target DSL.

For Comby, we further enhance the prompt using a Retrieval-Augmented Generation (RAG) strategy. Leveraging MELT \cite{ramos_melt_2023}, a tool for inferring transformation rules from pull requests, we collected over 1,700 Comby rules. At inference time, we retrieve the 8 most relevant examples to include in the prompt, providing higher-quality and more targeted guidance.

In contrast, we did not identify any equivalent large-scale rule repositories or RAG mechanisms for Ast-Grep and GritQL. Instead, their prompts include a fixed set of 8 examples \orange{derived from the DSL official documentation}. As a result, the RAG-based evaluation is conducted only for Comby, enabling us to assess the impact of retrieval augmentation separately from standard in-context learning.

%\orange{Prompt development followed an iterative refinement process to improve output quality and consistency across models. We evaluated both zero-shot and few-shot prompting strategies, with and without examples, and added structured output constraints (JSON, YAML) to reduce formatting errors.}

\camera{\textbf{As baseline}, we consider an anti-unification algorithm~\cite{ketkar_inferring_2022} that takes as input the pairs of code examples from our six datasets and identifies the most general structure shared by two expressions; in our case, the match pattern and the corresponding rewrite pattern.}

\removed{The rules are written in configuration files in YAML or TOML format depending on the DSL containing the name/id of the rule, the match pattern and the rewrite pattern.} We execute the transformation rules on the input/before version to obtain a transformed version of the code representing the output of the generated rule. Once the transformation rule is generated, we apply it to the original before-code snippet to produce the transformed code. Next, we compare the generated code with the ground-truth version in order to compute the evaluation metrics described earlier, including syntactic and semantic equivalence.

The evaluation process first assesses the \textit{applicability} of the generated rules, verifying that they conform to the syntax of the target DSL and can be successfully compiled and executed by the corresponding DSL engine. We also report the number of consumed tokens as a measure of computational cost. We then evaluate the ability of the LLMs to leverage the expressiveness of the DSLs, particularly through the use of metavariable placeholders that enable rule generalization. We also characterize the number and size of the generated rules. \camera{Generalizability is also assessed using a reuse score, computed by applying the rules derived from correct transformations to all instances in each dataset and counting the number of additional matches, i.e., how many times a rules matches.}

After that, we also measure the correctness of the produced transformations by comparing the generated code with the ground-truth version. This comparison includes both \textit{exact syntax matching} (considering formatting aspects such as spaces and indentation) and \textit{AST-level matching}, which relies on parsing the code into abstract syntax trees to assess structural equivalence. In addition, we compute the tree edit distance to quantify the structural differences between the generated and reference code.
\camera{We further strengthen semantic evaluation by leveraging the Defects4J and BugsInPy test suites to assess alignment between ground-truth and patched code.}

Additionally, to better understand and contextualize our results, we conducted a qualitative analysis for the first three research questions (RQs). Specifically, for each dataset and each DSL, we randomly selected a subset of 72 pairs while applying additional criteria tailored to each RQ. The random selection was performed by choosing one pair for each combination of benchmark × DSL × model (6 × 4 × 3 = 72). However, for RQ1, some DSL–model combinations exhibited 100\% RA, reducing the total number of selected pairs to 60.
For RQ1, we focused on cases where the generated rules were not applicable, i.e., violated the DSL syntax, or could not be executed by the transformation engine.
For RQ2, we selected cases where the generated rules were valid and executable, enabling us to analyze how LLMs leverage DSL expressiveness, particularly through metavariable usage and rule generalization.
For RQ3, we considered cases where the generated transformations were applicable but did not exactly match the ground truth, allowing us to analyze syntax-level discrepancies, such as incorrect, missing, or unneeded edits, and to identify notable patterns and outlier results.

Finally, to assess whether observed differences between models are  statistically significant, we apply McNemar's test for paired binary metrics (RA, EM, AM) and Wilcoxon signed-rank tests for the continuous metric (TD), with Benjamini-Hochberg correction for multiple comparisons
across the (6*4) 24 dataset--DSL configurations ($\alpha = 0.05$). For
aggregate-level metrics (\#R, SR, MTR), we use a sign test. Full results and scripts in the replication package.

\section{Results}

\subsection{RQ1: Effectiveness}%: {\small To what extent can LLMs generate sound transformation rules from code diffs while maintaining reasonable computational and financial cost?}}

\begin{table}[t]
\centering
\footnotesize
\caption{Soundness of generated transformation rules per dataset, tool, and LLM (grouped by scenario)}
\vspace{\vspacesarebad} % à modifier dans main.tex pour être consistant
\setlength{\tabcolsep}{3pt}
\label{tab:results-rq1-bis}
\begin{tabular}{l*{4}{c@{\hspace{2pt}}c@{\hspace{2pt}}c}}

\toprule
\textbf{Tool}
& \multicolumn{3}{c}{\textbf{GPT-oss-120B}}
& \multicolumn{3}{c}{\textbf{GPT5.4}}
& \multicolumn{3}{c}{\textbf{Llama3.1-8B}}
& \multicolumn{3}{c}{\camera{\textbf{Anti-Uni}}} \\
\cmidrule(lr){2-4}
\cmidrule(lr){5-7}
\cmidrule(lr){8-10}
\cmidrule(lr){11-13}
& \textbf{F} & \textbf{RA} & \textbf{NT}
& \textbf{F} & \textbf{RA} & \textbf{NT}
& \textbf{F} & \textbf{RA} & \textbf{NT}
& \camera{\textbf{F}} & \camera{\textbf{RA}} & \camera{\textbf{NT}} \\
\midrule

\multicolumn{13}{c}{\textbf{API Misuse} --- \emph{Galappaththi et al. (n=43)}}\\
\addlinespace[1pt]
Comby \textit{(R)}
& 3 & 93.0\% & 192
& 2 & 95.3\% & 208
& 18 & 58.1\% & \textbf{166}
& \textbf{1} & \textbf{97.6\%} & 283 \\
Comby \textit{(NR)}
& \textbf{0} & \textbf{100.0\%} & \textbf{185}
& \textbf{0} & \textbf{100.0\%} & 202
& 8 & 81.4\% & 264
& -- & -- & -- \\
Ast-Grep
& 17 & 60.5\% & \textbf{170}
& \textbf{3} & \textbf{93.0\%} & 228
& 25 & 40.5\% & 342
& 19 & 54.8\% & 254 \\
GritQL
& 2 & 95.3\% & 253
& \textbf{0} & \textbf{100.0\%} & \textbf{215}
& 1 & 97.7\% & 293
& \textbf{0} & \textbf{100.0\%} & 279 \\

\cmidrule(lr){1-13}

\multicolumn{13}{c}{\textbf{Program Repair} --- \emph{ManySStuBs4J (n=528)}}\\
\addlinespace[1pt]
Comby \textit{(R)}
& 9 & 98.9\% & \textbf{90}
& \textbf{5} & \textbf{99.2\%} & 225
& 335 & 36.6\% & 212
& \textbf{2} & \textbf{100.0\%} & 460 \\
Comby \textit{(NR)}
& \textbf{12} & \textbf{97.9\%} & \textbf{88}
& 88 & 83.5\% & 171
& 128 & 76.1\% & 995
& -- & -- & -- \\
Ast-Grep
& 219 & 58.7\% & 108
& \textbf{77} & \textbf{85.6\%} & \textbf{108}
& 448 & 15.5\% & 557
& 211 & 54.1\% & 451 \\
GritQL
& 5 & 99.4\% & \textbf{148}
& 5 & 99.4\% & 221
& \textbf{4} & \textbf{99.8\%} & 242
& \textbf{2} & \textbf{100.0\%} & 480 \\

\addlinespace[1pt]
\multicolumn{13}{c}{\textbf{Program Repair} --- \emph{Defects4J (n=427)}}\\
\addlinespace[1pt]
Comby \textit{(R)}
& 34 & 92.3\% & 160
& \textbf{4} & \textbf{99.1\%} & 211
& 264 & 38.0\% & \textbf{124}
& \textbf{0} & \textbf{100.0\%} & \textbf{82} \\
Comby \textit{(NR)}
& \textbf{24} & \textbf{94.8\%} & \textbf{134}
& 89 & 79.1\% & 173
& 125 & 71.1\% & 239
& -- & -- & -- \\
Ast-Grep
& 203 & 53.8\% & 152
& \textbf{92} & \textbf{78.4\%} & 179
& \textbf{30} & \textbf{95.1\%} & 262
& 299 & 29.8\% & \textbf{79} \\
GritQL
& 6 & 98.6\% & 201
& \textbf{0} & \textbf{100.0\%} & 240
& \textbf{0} & \textbf{100.0\%} & 174
& \textbf{0} & \textbf{100.0\%} & 100 \\

\addlinespace[1pt]
\multicolumn{13}{c}{\textbf{Program Repair} --- \emph{BugsInPy (n=559)}}\\
\addlinespace[1pt]
Comby \textit{(R)}
& 36 & 93.6\% & 158
& 34 & 93.9\% & 213
& 429 & 23.3\% & 159
& \textbf{16} & \textbf{97.1\%} & \textbf{156} \\
Comby \textit{(NR)}
& \textbf{18} & \textbf{96.8\%} & \textbf{149}
& 178 & 68.2\% & 169
& 134 & 76.0\% & 201
& -- & -- & -- \\
Ast-Grep
& 305 & 45.4\% & \textbf{140}
& \textbf{97} & \textbf{82.6\%} & 187
& 404 & 27.7\% & 336
& 401 & 28.0\% & 149 \\
GritQL
& 2 & 99.6\% & 212
& \textbf{0} & \textbf{100.0\%} & 248
& 1 & 99.8\% & 185
& \textbf{0} & \textbf{100.0\%} & \textbf{172} \\

\cmidrule(lr){1-13}

\multicolumn{13}{c}{\textbf{API Migration} --- \emph{PyMigBench (n=605)}}\\
\addlinespace[1pt]
Comby \textit{(R)}
& 55 & 90.9\% & 293
& 64 & 89.4\% & 394
& 367 & 39.3\% & \textbf{183}
& \textbf{42} & \textbf{93.0\%} & 421 \\
Comby \textit{(NR)}
& \textbf{40} & \textbf{93.4\%} & \textbf{241}
& 90 & 85.1\% & 323
& 190 & 68.6\% & 768
& -- & -- & -- \\
Ast-Grep
& 311 & 48.6\% & \textbf{183}
& \textbf{114} & \textbf{81.2\%} & 349
& 462 & 23.6\% & 399
& 565 & 6.5\% & 617 \\
GritQL
& 4 & 99.3\% & 365
& 1 & 99.8\% & 471
& 4 & 99.3\% & \textbf{255}
& \textbf{0} & \textbf{100.0\%} & 666 \\

\cmidrule(lr){1-13}

\multicolumn{13}{c}{\textbf{Language Version Migration} --- \emph{JMigBench (n=45)}}\\
\addlinespace[1pt]
Comby \textit{(R)}
& \textbf{0} & \textbf{100.0\%} & 113
& \textbf{0} & \textbf{100.0\%} & 141
& 20 & 55.6\% & \textbf{112}
& \textbf{0} & \textbf{100.0\%} & 141 \\
Comby \textit{(NR)}
& \textbf{0} & \textbf{100.0\%} & \textbf{113}
& 1 & 97.8\% & 146
& 6 & 86.7\% & 173
& -- & -- & -- \\
Ast-Grep
& 19 & 57.8\% & \textbf{119}
& \textbf{12} & \textbf{73.3\%} & 165
& 32 & 28.9\% & 198
& 34 & 24.4\% & 133 \\
GritQL
& \textbf{0} & \textbf{100.0\%} & \textbf{139}
& \textbf{0} & \textbf{100.0\%} & 161
& 1 & \textbf{100.0\%} & 161
& \textbf{0} & \textbf{100.0\%} & 156 \\

\bottomrule
\end{tabular}
\newline\textbf{\textit{F: Failures; RA: Rule Applicability rate; NT: Number of Generated Tokens; R: RAG; NR: no RAG.}}
\vspace{\vspacesarebad}
\end{table}
\autoref{tab:results-rq1-bis} reports the soundness of transformation rules generated by GPT-oss-120B, GPT-5.4, and Llama3.1-8B across all datasets and DSLs. We report on three metrics: the number of Failures (F), Rule Applicability (RA\%), and the Number of Tokens (NT).
Values in bold correspond to the top 1\% of results for each tool.

Overall, GPT-5.4 achieves the highest Rule Applicability (RA), with values ranging from 68.2\% to 100\%, and consistently records the lowest number of failures when using Ast-Grep, GritQL, and Comby with RAG. Across these settings, RA frequently exceeds 90\%, reaching up to 100\% in several cases. 
GritQL stands out as the most reliable DSL, achieving the best RA across all models and datasets, often reaching near-perfect or perfect soundness.
In contrast, GPT-oss-120B performs best with Comby without RAG, producing fewer failures than other models while maintaining competitive RA, ranging from 93.6\% to 100\%. In this setting, GPT-5.4 exhibits a notable drop in performance, reaching as low as 68.2\%. Additionally, GPT-oss-120B consistently generates more concise outputs, with token counts typically 20–40\% lower than GPT-5.4.

\camera{As for anti-unification, we observe a high Rule Applicability (RA) for Comby and GritQL (93\%-100\%), reflecting the deterministic nature of the algorithm. In contrast, RA is much lower for Ast-Grep (6.5\%-54.8\%) because the generalized match expressions produced often fail to satisfy its requirement that patterns should correspond to valid AST nodes, making many generated rules inapplicable.}

\emph{\textbf{Statistical significance.}} McNemar tests confirm these trends: GPT-5.4 is significantly better than GPT-oss-120B in 7 of 24 configurations ($p < 0.05$),
  mostly on Ast-Grep, while GPT-oss-120B wins in 4, all on Comby without
  RAG; both significantly outperform Llama3.1-8B in at least 16 of 24.

While GPT-5.4 delivers the strongest overall performance at the cost of increased token usage. Llama3.1-8B further amplifies this trend, often producing the largest outputs (e.g., exceeding 300 tokens) while also yielding lower RA and higher failure rates. 

A deeper qualitative analysis of the random sample of 60 cases reveals that GPT-5.4 generates more tokens primarily because it handles more complex transformations, particularly in datasets such as Defects4J, BugsInPy, and PyMigBench, which involve larger code changes. Indeed, as shown in Table 1, these benchmarks involve function-level or multi-statement bug fixes, requiring larger and more complex transformation rules that GPT-oss-120B and Llama3.1-8B often fail to generate. In contrast, for simpler transformations, GPT-5.4 produces a number of tokens comparable to GPT-oss-120B. 
Moreover, most failures stem from formatting issues in the generated configurations, such as the inclusion of generated comments or extra formatting artifacts (e.g., inserting natural language explanations like \textit{// this rule replaces X with Y}, which break the DSL syntax). We also observe frequent errors related to missing rule separators in Ast-Grep, where multiple rules are written consecutively without proper delimiters\reviewonly{ (e.g., \texttt{rule1 {}-{}-{}- rule2})}, leading to parsing failures. Finally, we identify invalid patterns involving multiple matching nodes, such as attempting to match two unrelated AST nodes within a single rule, which is not supported by Ast-Grep and GritQL.

\begin{tcolorbox}
\textbf{$\boldsymbol{RQ_1}$ insights and findings:}
%GPT-5.4 achieves the best overall performance, with the highest RA (68.2\%–100\%) and the lowest number of failures across most DSLs. GritQL consistently delivers the highest soundness across all models.
%
%However, GPT-5.4 generates significantly more tokens, largely due to its ability to handle more complex transformations. In contrast, GPT-oss-120B achieves strong performance on Comby without RAG, with RA ranging from 93.6\% to 100\%.
%
%Llama3.1-8B underperforms overall, exhibiting lower RA and higher failure rates.
%
GPT-5.4 achieves the best overall soundness, with the highest rule applicability (68.2\%--100\%) and the fewest failures across most DSLs (GritQL being the most reliable DSL). However, it generates significantly more tokens, largely due to its ability to handle more complex transformations. GPT-oss-120B achieves strong performance specifically on Comby without RAG (93.6\%--100\%), while Llama3.1-8B exhibits lower RA and higher failure rates. \camera{Anti-unification achieves near-perfect RA on Comby and GritQL but low RA on Ast-Grep due to its stricter AST node constraints.}
%\textit{Takeaway:} 
 %The main finding is that rule synthesis is no longer limited by whether LLMs can capture transformation logic; the main barrier to soundness is compliance with DSL syntax and configuration conventions (e.g., rule separators, valid patterns, YAML/TOML structure).
 Overall, strong LLMs can generate executable transformation rules at a high rate. %Most 
 Remaining failures stem from DSL syntax and  %configuration 
 conventions rather than incorrect match and rewrite code. %from an inability to produce match and rewrite patterns. % (e.g., rule separators, valid patterns, YAML/TOML structure)
\end{tcolorbox}

\subsection{RQ2: Rule Quality}%: {\small How complete are the generated transformation rules in terms of coverage, structural richness, and generalization capacity?}}

\begin{table}[t]
\centering
\footnotesize
\caption{Soundness of generated transformation rules per Dataset, Tool, and LLM (grouped by scenario)}
\vspace{\vspacesarebad} % à modifier dans main.tex pour être consistant
\setlength{\tabcolsep}{3pt}
\label{tab:results-rq2-anti}
\begin{tabular}{l*{4}{c@{\hspace{2pt}}c@{\hspace{2pt}}c}}
\toprule
\textbf{Tool}
& \multicolumn{3}{c}{\textbf{GPT-oss-120B}}
& \multicolumn{3}{c}{\textbf{GPT5.4}}
& \multicolumn{3}{c}{\textbf{Llama3.1-8B}}
& \multicolumn{3}{c}{\camera{\textbf{Anti-Uni}}} \\ 
\cmidrule(lr){2-4}
\cmidrule(lr){5-7}
\cmidrule(lr){8-10}
\cmidrule(lr){11-13}
& \textbf{NR} & \textbf{\#R} & \textbf{MTR}
& \textbf{NR} & \textbf{\#R} & \textbf{MTR}
& \textbf{NR} & \textbf{\#R} & \textbf{MTR}
& \camera{\textbf{NR}} & \camera{\textbf{\#R}} & \camera{\textbf{MTR}} \\
\midrule

\multicolumn{13}{c}{\textbf{API Misuse} --- \emph{Galappaththi et al. (n=43)}}\\
\addlinespace[1pt]
Comby \textit{(R)}
& 3.4 & 57.3 & 7
& 1.6 & 162.8 & 10
& 3.9 & 44.5 & 14
& 1.9 & 113.7 & 54 \\
Comby \textit{(NR)}
& 3.4 & 53.9 & 7
& 2.1 & 115.3 & 6
& 7.3 & 41.4 & 3
& -- & -- & -- \\
Ast-Grep
& 3.1 & 39.6 & 17
& 3.6 & 51.0 & 9
& 4.3 & 48.3 & 15
& 1.9 & 96.7 & 53 \\
GritQL
& 2.8 & 69.6 & 7
& 1.9 & 93.8 & 13
& 4.2 & 46.9 & 0
& 1.9 & 96.7 & 37 \\

\cmidrule(lr){1-13}
\multicolumn{13}{c}{\textbf{Program Repair} --- \emph{ManySStuBs4J (n=528)}}\\
\addlinespace[1pt]
Comby \textit{(R)}
& 1.8 & 63.4 & 18
& 1.6 & 167.6 & 10
& 6.1 & 48.9 & 4
& 2.4 & 165.5 & 62 \\
Comby \textit{(NR)}
& 1.9 & 63.5 & 17
& 1.9 & 73.8 & 6
& 29.4 & 43.7 & 18
& -- & -- & -- \\
Ast-Grep
& 2.0 & 55.1 & 20
& 1.8 & 58.6 & 19
& 11.1 & 60.9 & 6
& 2.4 & 156.7 & 62 \\
GritQL
& 1.9 & 65.9 & 14
& 1.6 & 79.5 & 16
& 3.5 & 54.3 & 1
& 2.4 & 156.6 & 44 \\

\addlinespace[1pt]
\multicolumn{13}{c}{\textbf{Program Repair} --- \emph{Defects4J (n=427)}}\\
\addlinespace[1pt]
Comby \textit{(R)}
& 1.8 & 96.2 & 13
& 1.6 & 183.8 & 7
& 3.9 & 45.2 & 4
& 1.2 & 57.2 & 56 \\
Comby \textit{(NR)}
& 1.9 & 67.5 & 16
& 1.6 & 160.1 & 3
& 9.5 & 27.1 & 1
& -- & -- & -- \\
Ast-Grep
& 2.1 & 51.6 & 19
& 2.0 & 78.1 & 14
& 8.0 & 56.6 & 5
& 1.2 & 49.3 & 55 \\
GritQL
& 1.9 & 78.5 & 15
& 1.7 & 132.6 & 11
& 2.4 & 48.7 & 0
& 1.2 & 49.3 & 39 \\

\addlinespace[1pt]
\multicolumn{13}{c}{\textbf{Program Repair} --- \emph{BugsInPy (n=559)}}\\
\addlinespace[1pt]
Comby \textit{(R)}
& 2.1 & 80.8 & 15
& 1.7 & 186.2 & 4
& 4.4 & 42.4 & 6
& 1.7 & 74.1 & 61 \\
Comby \textit{(NR)}
& 2.1 & 75.9 & 13
& 1.7 & 147.4 & 3
& 5.8 & 43.6 & 2
& -- & -- & -- \\
Ast-Grep
& 2.0 & 54.8 & 18
& 2.3 & 76.7 & 11
& 6.7 & 51.3 & 8
& 1.7 & 63.3 & 60 \\
GritQL
& 2.1 & 77.8 & 13
& 1.9 & 131.3 & 8
& 2.8 & 42.7 & 1
& 1.7 & 59.1 & 43 \\

\cmidrule(lr){1-13}
\multicolumn{13}{c}{\textbf{API Migration} --- \emph{PyMigBench (n=605)}}\\
\addlinespace[1pt]
Comby \textit{(R)}
& 4.7 & 90.4 & 7
& 3.0 & 242.1 & 3
& 5.9 & 39.6 & 4
& 4.8 & 96.2 & 49 \\
Comby \textit{(NR)}
& 4.6 & 68.8 & 9
& 3.6 & 160.4 & 3
& 24.5 & 35.4 & 1
& -- & -- & -- \\
Ast-Grep
& 3.3 & 36.3 & 14
& 5.4 & 70.2 & 8
& 8.0 & 52.5 & 4
& 5.5 & 128.5 & 47 \\
GritQL
& 4.4 & 73.0 & 7
& 3.7 & 145.5 & 7
& 4.1 & 39.1 & 0
& 5.5 & 124.1 & 34 \\

\cmidrule(lr){1-13}
\multicolumn{13}{c}{\textbf{Language Version Migration} --- \emph{JMigBench (n=45)}}\\
\addlinespace[1pt]
Comby \textit{(R)}
& 1.8 & 115.8 & 17
& 1.1 & 278.6 & 13
& 3.2 & 45.8 & 10
& 1.4 & 144.1 & 31 \\
Comby \textit{(NR)}
& 1.6 & 135.1 & 14
& 1.1 & 285.1 & 7
& 5.7 & 41.9 & 7
& -- & -- & -- \\
Ast-Grep
& 2.0 & 60.0 & 20
& 2.9 & 72.1 & 22
& 4.7 & 47.2 & 11
& 1.4 & 135.0 & 29 \\
GritQL
& 1.3 & 149.8 & 16
& 1.2 & 204.7 & 13
& 2.6 & 53.2 & 0
& 1.4 & 135.0 & 20 \\

\bottomrule
\end{tabular}
\newline\textbf{\textit{\#R: number of Rules; SR: Size of Rules; MTR: Metavariables Token Ratio; R: RAG; NR: no RAG.}}
\vspace{\vspacesarebad}
\end{table}
\autoref{tab:results-rq2-anti} summarizes the structural completeness of transformation rules generated by GPT-oss-120B, GPT-5.4 and Llama3.1-8B across all datasets and DSLs using the metrics number of Rules (\#R), Size of Rules (SR), and the Metavariable Token Ratio (MTR\%).

Overall, a clear trend emerges across all benchmarks. GPT-5.4 consistently produces larger rules, with the highest SR values (e.g., up to 285.1 tokens on JMigBench), while generating fewer rules (\#R typically between 1.1 and 3.7) and using fewer metavariables (MTR mostly between 3\% and 13\%). In contrast, GPT-oss-120B generates more compact rules, with slightly higher \#R (typically 1.6 to 4.7) and significantly higher MTR (ranging from 7\% to 20\%).
This difference is particularly visible in program repair benchmarks such as Defects4J and BugsInPy, where GPT-oss-120B reaches MTR values up to 19–20\% with \#R around 2.0, while GPT-5.4 produces fewer rules (\#R = 1.6–2.3) with lower MTR (as low as 3–11\%). Similarly, in PyMigBench, GPT-5.4 generates larger rules (SR = 242.1) with low MTR (3\%), whereas GPT-oss-120B maintains higher values (MTR = 7–14\%) with slightly higher \#R (= 3.3–4.7).
This highlights a trade-off in rule construction strategies. GPT-5.4 tends to generate monolithic and explicit transformation rules, favoring larger patterns with limited abstraction. Conversely, GPT-oss-120B produces more modular and generalized rules, leveraging metavariables more extensively. 
Llama3.1-8B follows a similar pattern to GPT-oss-120B in terms of rule size but exhibits lower structural completeness overall, with fewer rules and limited use of metavariables.

\camera{Anti-unification consistently achieves the highest MTR across all instances because it transforms every common token between the match and rewrite patterns into metavariables. As a result, it reaches a maximum MTR of 62, compared to only 20 for all LLM-generated rules. This over-generalization leads to overly generic and unreadable rules that can match nearly any code fragment, highlighting the limitations of anti-unification when task-specific constraints are not implemented.}

\emph{\textbf{Statistical significance.}} A sign test across the 24 configurations confirms that GPT-5.4 produces larger rules (higher SR in 24/24, $p < 10^{-6}$) with fewer metavariables (lower MTR in 19/24, $p = 0.007$) than GPT-oss-120B, while Llama3.1-8B   generates more rules than both in 22/24 ($p < 10^{-4}$).

Moreover, our qualitative analysis on the random sample of 72 cases provides further insight into these trends. For complex transformations, particularly in datasets such as Defects4J, BugsInPy, and PyMigBench, GPT-5.4 demonstrates a stronger ability to leverage DSL expressiveness. For instance, it frequently uses metavariables such as \texttt{\$\$\$BODY} to capture entire sequences of AST nodes, enabling it to rewrite or remove complete function or method bodies in a single rule. Overall, it produces semantically meaningful metavariable names,
% such as \texttt{buf} for buffers, \texttt{col} for columns, or \texttt{beforeloop} when targeting code regions preceding a loop,
greatly improving readability for the generated rules.

In contrast, smaller models such as GPT-oss-120B and Llama3.1-8B tend to avoid large or complex transformations, focusing instead on simpler, localized edits. This often results in partial rules that fail to capture substantial code modifications. Additionally, GPT-oss-120B frequently introduces metavariables such as \texttt{:[indent]} to explicitly match whitespace, compensating for the lack of native indentation handling in Comby. While this allows for exact matching \removed{as required by the prompt,} it significantly increases rule verbosity and reduces generalization.

\begin{table}[t]
\centering
\footnotesize
\ifBlueHighlights\color{blue}\fi
\caption{\camera{Reuse statistics per dataset, DSL, and LLM (grouped by scenario)}}
\vspace{\vspacesarebad} % à modifier dans main.tex pour être consistant
\setlength{\tabcolsep}{4pt}
\label{tab:reuse}
\begin{tabular}{l*{3}{ccc}}
\toprule
\textbf{DSL}
& \multicolumn{3}{c}{\textbf{GPT-oss-120B}}
& \multicolumn{3}{c}{\textbf{GPT5.4}}
& \multicolumn{3}{c}{\textbf{Llama3.1-8B}} \\
\cmidrule(lr){2-4}
\cmidrule(lr){5-7}
\cmidrule(lr){8-10}
& \textbf{Med.} & \textbf{2+} & \textbf{20+}
& \textbf{Med.} & \textbf{2+} & \textbf{20+}
& \textbf{Med.} & \textbf{2+} & \textbf{20+} \\
\midrule

\multicolumn{10}{c}{\textbf{API Misuse} --- \emph{Galappaththi et al. (n=43)}}\\
\addlinespace[1pt]
Comby \textit{(R)}
& 1.0 & 10 & 1 & 1.0 & 5 & 0 & 1.0 & 2 & 0 \\
Comby \textit{(NR)}
& 1.0 & \textbf{11} & \textbf{3} & 1.0 & 3 & 1 & 1.0 & 3 & \textbf{2} \\
Ast-Grep
& \textbf{1.5} & 6 & 2 & \textbf{3.0} & \textbf{15} & \textbf{4} & \textbf{3.0} & 2 & 0 \\
GritQL
& 1.0 & 6 & 2 & 1.0 & 5 & 2 & 1.0 & \textbf{4} & \textbf{2} \\

\cmidrule(lr){1-10}
\multicolumn{10}{c}{\textbf{Program Repair} --- \emph{ManySStuBs4J (n=528)}}\\
\addlinespace[1pt]
Comby \textit{(R)}
& 1.0 & 84 & 32 & \textbf{1.0} & 63 & 11 & 0.0 & 35 & 15 \\
Comby \textit{(NR)}
& 1.0 & 127 & 35 & \textbf{1.0} & 91 & 17 & 0.0 & \textbf{793} & \textbf{100} \\
Ast-Grep
& \textbf{2.0} & 151 & 42 & \textbf{1.0} & \textbf{220} & \textbf{64} & \textbf{1.0} & 21 & 6 \\
GritQL
& 1.0 & \textbf{166} & \textbf{44} & \textbf{1.0} & 98 & 48 & 0.0 & 59 & 14 \\

\addlinespace[1pt]
\multicolumn{10}{c}{\textbf{Program Repair} --- \emph{Defects4J (n=427)}}\\
\addlinespace[1pt]
Comby \textit{(R)}
& 1.0 & 33 & 15 & 1.0 & 10 & 3 & 1.0 & 5 & 5 \\
Comby \textit{(NR)}
& 1.0 & 45 & \textbf{18} & 1.0 & 12 & 7 & 1.0 & 16 & \textbf{8} \\
Ast-Grep
& 2.0 & 42 & 15 & 1.0 & 63 & \textbf{21} & 0.0 & 1 & 1 \\
GritQL
& \textbf{3.0} & \textbf{51} & 4 & \textbf{3.0} & \textbf{212} & 8 & \textbf{3.0} & \textbf{45} & 6 \\

\addlinespace[1pt]
\multicolumn{10}{c}{\textbf{Program Repair} --- \emph{BugsInPy (n=559)}}\\
\addlinespace[1pt]
Comby \textit{(R)}
& 1.0 & \textbf{56} & \textbf{15} & 1.0 & 19 & 3 & 1.0 & 9 & \textbf{5} \\
Comby \textit{(NR)}
& 1.0 & \textbf{56} & 14 & 1.0 & 34 & 4 & 1.0 & \textbf{31} & \textbf{5} \\
Ast-Grep
& \textbf{2.5} & 40 & 6 & \textbf{2.0} & \textbf{118} & \textbf{17} & \textbf{5.0} & 6 & 1 \\
GritQL
& 1.0 & 29 & 3 & 1.0 & 67 & 6 & 1.0 & 15 & 1 \\

\cmidrule(lr){1-10}
\multicolumn{10}{c}{\textbf{API Migration} --- \emph{PyMigBench (n=605)}}\\
\addlinespace[1pt]
Comby \textit{(R)}
& 4.0 & 281 & 102 & 2.0 & 176 & 36 & 3.0 & 40 & 5 \\
Comby \textit{(NR)}
& 6.0 & 288 & \textbf{111} & 3.0 & 242 & 41 & 6.0 & \textbf{273} & \textbf{44} \\
Ast-Grep
& \textbf{13.5} & 115 & 51 & 10.0 & 245 & \textbf{112} & \textbf{33.5} & 10 & 8 \\
GritQL
& 12.0 & \textbf{379} & 70 & \textbf{12.0} & \textbf{478} & 75 & 10.0 & 180 & 31 \\

\cmidrule(lr){1-10}
\multicolumn{10}{c}{\textbf{Language Version Migration} --- \emph{JMigBench (n=45)}}\\
\addlinespace[1pt]
Comby \textit{(R)}
& 1.0 & 6 & 0 & 1.0 & 1 & 0 & 1.0 & \textbf{3} & 0 \\
Comby \textit{(NR)}
& 1.0 & 1 & 0 & 1.0 & 0 & 0 & 1.0 & 2 & 0 \\
Ast-Grep
& 1.5 & 1 & 0 & 2.0 & 7 & 1 & 0.0 & 0 & 0 \\
GritQL
& \textbf{91.0} & \textbf{12} & \textbf{12} & \textbf{91.0} & \textbf{16} & \textbf{16} & \textbf{91.0} & 2 & \textbf{2} \\

\bottomrule
\end{tabular}
\newline
\textbf{\textit{Med.: Median matches per rule; $k+$: number of rules with more than $k$ matches; R: RAG; NR: no RAG.}}
\vspace{\vspacesarebad}
\end{table}

\camera{\emph{\textbf{Generalizability beyond a single example.}}}
\camera{\autoref{tab:reuse} presents the results of the reusability experiments.
For GPT-oss-120B, GritQL produced the largest number of reusable rules, with 166 rules matching at least twice and 44 matching more than 20 times, closely followed by Ast-Grep (151 and 42, respectively). GPT5.4 exhibits a similar trend, with Ast-Grep yielding 220 rules reused at least twice and 64 reused more than 20 times, compared with 98 and 48 for GritQL. In contrast, Comby generated fewer highly reusable rules.}
\camera{Llama3.1-8B consistently achieves the lowest reuse across most DSLs, typically with fewer than 60 rules reused at least twice. The main exception is Comby without RAG, which generated 793 rules reused at least twice and 100 reused more than 20 times. However, these rules have a median reuse of 0, indicating that this behavior is driven by a relatively small number of highly repetitive transformations. A manual inspection revealed many overly generic rules, explaining why this configuration achieves lower AM/EM scores despite producing a large number of reusable matches.}

\camera{In addition, we provide more detailed illustrations of the reuse score for each dataset by plotting the most frequently matched rules for each DSL and model on a logarithmic scale. These figures better show the high reuse score in the first quartile and the outliers with a very high number of matches, corresponding to recurrent transformations such as \texttt{assert(:[var])} $\rightarrow$ \texttt{Assert.assert(:[var])}. They are available in the \href{https://anonymous.4open.science/r/EmpiricalStudyTransformationRules-13B3/evaluation/rule_reuse/results/}{replication package [link]}.}

\begin{tcolorbox}
\textbf{$\boldsymbol{RQ_2}$ insights and findings:}
%
%GPT-5.4 generates larger, less abstract rules (high SR, low MTR, low \#R), while GPT-oss-120B produces more compact and generalized rules (higher MTR and \#R). 
%
%This reflects a trade-off: GPT-5.4 better handles complex transformations with explicit rules, whereas smaller models focus on simpler changes and often miss larger modifications.
%
GPT-5.4 generates larger, more explicit rules (high SR, low MTR, low \#R), while GPT-oss-120B produces more compact and generalized rules (higher MTR and \#R). \camera{In contrast, the anti-unification baseline achieves the highest MTR by over-generalizing transformation rules.} Smaller models focus on simpler changes and often miss larger transformations.
The qualitative analysis reflects a trade-off between abstraction and coverage. More abstract, metavariable-heavy rules are not automatically better, and explicitness is sometimes the mechanism enabling complex transformations to be captured at all.
\camera{The reuse score analysis further shows that many LLM-generated rules generalize beyond a single example.}
\end{tcolorbox}

\subsection{RQ3: Accuracy}%: {\small How well do the generated transformation rules reproduce the ground-truth transformations in terms of semantic and syntactic similarity?}}

\begin{table}[t]
\centering
\footnotesize
\caption{Syntactic\removed{and semantic} alignment of generated code per dataset, tool, and LLM (grouped by scenario)}
\vspace{\vspacesarebad} % à modifier dans main.tex pour être consistant
\setlength{\tabcolsep}{4.5pt}
\label{tab:results-rq3-bis}
\begin{tabular}{l*{4}{c@{\hspace{2pt}}c@{\hspace{2pt}}c}}
\toprule
\textbf{Tool}
& \multicolumn{3}{c}{\textbf{GPT-oss-120B}}
& \multicolumn{3}{c}{\textbf{GPT5.4}}
& \multicolumn{3}{c}{\textbf{Llama3.1-8B}}
& \multicolumn{3}{c}{\camera{\textbf{Anti-Uni}}} \\
\cmidrule(lr){2-4}
\cmidrule(lr){5-7}
\cmidrule(lr){8-10}
\cmidrule(lr){11-13}
& \textbf{TD} & \textbf{EM} & \textbf{AM}
& \textbf{TD} & \textbf{EM} & \textbf{AM}
& \textbf{TD} & \textbf{EM} & \textbf{AM}
& \camera{\textbf{TD}} & \camera{\textbf{EM}} & \camera{\textbf{AM}} \\
\midrule

\multicolumn{13}{c}{\textbf{API Misuse} --- \emph{Galappaththi et al. (n=43)}}\\
\addlinespace[1pt]
Comby \textit{(R)}
& 82.8 & \textbf{67.5} & \textbf{67.5}
& \textbf{88.1} & 63.4 & 63.4
& 80.1 & 24.0 & 24.0
& 71.5 & 0.0 & 0.0 \\
Comby \textit{(NR)}
& 79.7 & \textbf{53.5} & \textbf{53.5}
& \textbf{82.1} & 48.8 & 48.8
& 72.9 & 25.7 & 25.7
& -- & -- & -- \\
Ast-Grep
& 85.6 & 53.8 & 53.8
& \textbf{88.6} & \textbf{67.5} & \textbf{67.5}
& 83.1 & 17.6 & 17.6
& 65.7 & 17.4 & 17.4 \\
GritQL
& 86.3 & \textbf{58.5} & \textbf{58.5}
& \textbf{88.1} & 46.5 & 46.5
& 77.6 & 23.8 & 23.8
& 73.4 & 9.5 & 9.5 \\

\cmidrule(lr){1-13}

\multicolumn{13}{c}{\textbf{Program Repair} --- \emph{ManySStuBs4J (n=528)}}\\
\addlinespace[1pt]
Comby \textit{(R)}
& 96.7 & 45.1 & 45.1
& \textbf{97.0} & \textbf{57.6} & \textbf{57.6}
& 83.6 & 11.9 & 11.9
& 96.3 & 3.9 & 3.9 \\
Comby \textit{(NR)}
& 97.3 & 45.7 & 45.7
& \textbf{97.7} & \textbf{56.4} & \textbf{56.4}
& 89.5 & 18.2 & 18.2
& -- & -- & -- \\
Ast-Grep
& 90.8 & 52.1 & 52.1
& \textbf{94.3} & \textbf{62.5} & \textbf{62.5}
& 68.6 & 23.8 & 23.8
& 90.8 & 3.2 & 3.2 \\
GritQL
& \textbf{98.5} & 46.1 & 46.1
& 98.1 & \textbf{56.4} & \textbf{56.4}
& 93.9 & 5.3 & 5.3
& 95.3 & 1.5 & 1.5 \\

\addlinespace[1pt]
\multicolumn{13}{c}{\textbf{Program Repair} --- \emph{Defects4J (n=427)}}\\
\addlinespace[1pt]
Comby \textit{(R)}
& 92.5 & 31.6 & 31.6
& \textbf{94.1} & \textbf{33.6} & \textbf{33.6}
& 84.3 & 8.0 & 8.0
& 91.0 & 1.2 & 1.2 \\
Comby \textit{(NR)}
& 89.4 & 28.1 & 28.1
& \textbf{94.0} & \textbf{35.0} & \textbf{35.0}
& 84.0 & 9.3 & 9.3
& -- & -- & -- \\
Ast-Grep
& 89.5 & 41.3 & 41.3
& \textbf{90.0} & \textbf{49.4} & \textbf{49.4}
& 69.7 & 5.6 & 5.6
& 86.9 & 2.4 & 2.4 \\
GritQL
& 92.2 & 7.4 & 7.4
& \textbf{93.6} & \textbf{35.4} & \textbf{35.4}
& 90.7 & 8.5 & 8.5
& 89.4 & 1.4 & 1.4 \\

\addlinespace[1pt]
\multicolumn{13}{c}{\textbf{Program Repair} --- \emph{BugsInPy (n=493)}}\\
\addlinespace[1pt]
Comby \textit{(R)}
& 74.3 & \textbf{22.4} & \textbf{22.4}
& 54.9 & 11.8 & 11.8
& 71.8 & 6.9 & 6.9
& \textbf{87.2} & 3.0 & 3.0 \\
Comby \textit{(NR)}
& 67.3 & 19.0 & 19.0
& 74.5 & \textbf{28.9} & \textbf{28.9}
& \textbf{78.6} & 9.6 & 9.6
& -- & -- & -- \\
Ast-Grep
& 84.7 & 31.5 & 31.5
& \textbf{93.7} & \textbf{54.1} & \textbf{54.1}
& 68.3 & 5.8 & 5.8
& 78.1 & 15.4 & 15.4 \\
GritQL
& 84.0 & 18.0 & 18.0
& \textbf{92.9} & \textbf{33.5} & \textbf{33.5}
& 86.0 & 9.9 & 9.9
& 85.6 & 5.0 & 5.0 \\

\cmidrule(lr){1-13}

\multicolumn{13}{c}{\textbf{API Migration} --- \emph{PyMigBench (n=661)}}\\
\addlinespace[1pt]
Comby \textit{(R)}
& 69.2 & 43.3 & 43.3
& \textbf{69.6} & \textbf{45.7} & \textbf{45.7}
& 67.6 & 12.2 & 12.2
& 61.1 & 15.8 & 15.8 \\
Comby \textit{(NR)}
& 66.0 & 41.8 & 41.8
& 69.5 & \textbf{49.9} & \textbf{49.9}
& \textbf{72.1} & 25.5 & 25.5
& -- & -- & -- \\
Ast-Grep
& 88.3 & 46.3 & 46.3
& \textbf{89.3} & \textbf{57.6} & \textbf{57.6}
& 66.9 & 8.4 & 8.4
& 87.2 & 0.0 & 0.0 \\
GritQL
& 78.2 & 35.3 & 35.3
& \textbf{80.8} & \textbf{40.1} & \textbf{40.1}
& 76.2 & 17.3 & 17.3
& 75.5 & 0.0 & 0.0 \\

\cmidrule(lr){1-13}

\multicolumn{13}{c}{\textbf{Language Version Migration} --- \emph{JMigBench (n=45)}}\\
\addlinespace[1pt]
Comby \textit{(R)}
& \textbf{91.0} & \textbf{68.9} & \textbf{68.9}
& 87.0 & 60.0 & 60.0
& 67.8 & 20.0 & 20.0
& 76.2 & 8.9 & 8.9 \\
Comby \textit{(NR)}
& \textbf{88.8} & \textbf{62.2} & \textbf{62.2}
& 81.0 & 45.5 & 45.5
& 58.0 & 23.1 & 23.1
& -- & -- & -- \\
Ast-Grep
& 81.6 & 30.8 & 30.8
& 87.7 & \textbf{51.5} & \textbf{51.5}
& 55.7 & 0.0 & 0.0
& \textbf{98.8} & 36.4 & 36.4 \\
GritQL
& \textbf{76.9} & 22.2 & 22.2
& 76.9 & \textbf{24.4} & \textbf{24.4}
& 74.8 & 4.5 & 4.5
& 74.5 & 8.9 & 8.9 \\

\bottomrule
\end{tabular}
\newline\textbf{\textit{TD: Tree Distance; EM/AM: Exact/AST Match rate; R: RAG; NR: no RAG.}}
\vspace{\vspacesarebad}
\end{table}

\autoref{tab:results-rq3-bis} summarizes the syntactic and semantic alignment of transformations made through the rules generated by GPT-oss-120B, GPT-5.4, and Llama3.1-8B across six benchmarks, evaluated in terms of semantic and syntactic similarity using the metrics of Exact Match (EM), AST Match (AM), and Tree Distance (TD). Values highlighted in bold correspond to the top 1\% of results for each tool. 
Overall, GPT-5.4 achieves the best performance across all metrics, consistently outperforming other models on most benchmarks. It reaches the highest TD scores (up to 98.1\%) and strong EM/AM results, particularly on complex datasets such as Defects4J (up to 56.0\% AM) and BugsInPy (up to 54.1\% EM/AM). These results indicate that GPT-5.4 generates transformations that are both structurally and semantically closest to the ground truth. 
GPT-oss-120B performs competitively, especially with Comby, both with and without RAG. For instance, it achieves up to 67.5\% EM/AM on API misuse with Comby + RAG and maintains strong performance on JMigBench (up to 68.9\% EM). However, its performance drops on more complex transformations, where it struggles to fully capture large edits. 
In contrast, Llama3.1-8B consistently exhibits low performance across all benchmarks and DSLs, with EM/AM scores often below 30\% and dropping as low as 5.2\% on Defects4J, highlighting its limited ability to generate accurate transformations.

\camera{The anti-unification algorithm performs significantly worse, achieving at most 36.4\% EM/AM in the best-case scenario and only 1.2–2.4\% EM/AM on benchmarks such as Defects4J. These results primarily come from the algorithm’s simplistic, non-contextual understanding of code, and its inability to handle code additions.}

\emph{\textbf{Statistical significance.}} McNemar tests confirm that GPT-5.4 significantly outperforms GPT-oss-120B on AST Match in 12 of 24 configurations vs.\ only 2 wins for GPT-oss-120B, and outperforms Llama3.1-8B in all 24 (all $p < 0.05$); Wilcoxon tests on tree distance yield consistent results (13/24 significant).

Finally, our qualitative analysis on the random sample of 72 cases provides deeper insights by categorizing errors into edits, deletions, and additions. We observe that Llama3.1-8B frequently produces missing edits and incorrect transformations, and often generates duplicate rules with identical match and rewrite patterns, indicating limited ability to consolidate transformations and generalize across similar cases. This behavior contributes to its low overall correctness. 
GPT-oss-120B performs better but still exhibits missing edits, particularly in cases where the modification does not correspond to a complete AST node sequence. In such scenarios, the model should generalize the transformation by expanding the matching pattern to capture the full intended change. However, it often fails to do so, producing rules that mirror only a narrowly scoped pattern rather than a generalized one, leading to incomplete edits. 
In contrast, GPT-5.4 produces a higher proportion of correct edits, especially for complex and non-local transformations. % such as large configuration or CLI parser modifications.

We also evaluated the semantic equivalence of transformations that did not result in an AST match in our sample of 72 rules. 
On this subset, 36 transformations failed to produce an AST match, with only one instance preserving semantic equivalence. Overall, none of the transformations that resulted in differing edits achieved full semantic equivalence. This likely because the goal was to regenerate the same V2 of the code, i.e., exact syntactic equivalence.  %This outcome can largely be attributed to our prompt design, which explicitly instructs the model to generate syntactically identical code in accordance with the transformation rules.

The single case of preserved semantic equivalence involved a trivial pattern with a redundant variable declaration (e.g., \texttt{int x = 0; int x = 0; return x}). However, such instances do not constitute meaningful or practically useful transformations. Furthermore, some transformations may preserve semantic equivalence at a localized level (e.g., method, import, or class), while still producing code that is invalid or inconsistent at the file level.

\begin{table}[t]
\centering
\footnotesize
\ifBlueHighlights\color{blue}\fi
\setlength{\tabcolsep}{4pt}
\caption{\camera{Test validation on Defects4J and BugsInPy test suites}}
\label{tab:test-validation}
\begin{tabular}{
ll
*3{ccc}
}
\toprule
\textbf{Dataset} & \textbf{DSL}
& \multicolumn{3}{c}{\textbf{GPT-oss-120B}}
& \multicolumn{3}{c}{\textbf{GPT5.4}}
& \multicolumn{3}{c}{\textbf{Llama3.1-8B}} \\
\cmidrule(lr){3-5}
\cmidrule(lr){6-8}
\cmidrule(lr){9-11}
&
& \textbf{A} & \textbf{P} & \textbf{SR}
& \textbf{A} & \textbf{P} & \textbf{SR}
& \textbf{A} & \textbf{P} & \textbf{SR} \\
\midrule

\multirow{4}{*}{Defects4J}
& Comby (R) & 217 & 6 & 2.8 & 243 & \textbf{7} & \textbf{2.9} & 93 & 2 & 2.2 \\

& Comby (NR) & 227 & 5 & 2.2 & 24 & 0 & 0.0 & 183 & \textbf{2} & \textbf{1.1} \\

& Ast-Grep & 89 & \textbf{6} & \textbf{6.7} & 40 & 0 & 0.0 & 105 & 5 & 4.8 \\

& GritQL & 269 & \textbf{8} & \textbf{3.0} & 204 & 6 & 2.9 & 98 & 0 & 0.0 \\
\midrule

\multirow{4}{*}{BugsInPy}
& Comby (R) & 405 & \textbf{11} & 2.7 & 462 & 2 & 0.4 & 120 & 6 & \textbf{5.0} \\

& Comby (NR) & 437 & 18 & 4.1 & 270 & 8 & 3.0 & 383 & \textbf{23} & \textbf{6.0} \\

& Ast-Grep & 174 & 28 & 16.1 & 212 & \textbf{38} & \textbf{17.9} & 146 & 4 & 2.7 \\

& GritQL & 456 & 16 & 3.5 & 371 & \textbf{17} & \textbf{4.6} & 504 & 12 & 2.4 \\

\bottomrule
\end{tabular}
\textbf{\textit{A/P: Attempted/Successful Repairs; SR: Success Rate (\%); R: RAG; NR: no RAG.}}

\end{table}

\begin{figure}[t]
\centering
\begin{tikzpicture}

% Circles
\draw[draw=darkgray, fill=red!20,   opacity=0.5] (-0.5,0.0) circle (1.0);
\draw[draw=darkgray, fill=green!20, opacity=0.5] ( 0.5,0.0) circle (1.0);
\draw[draw=darkgray, fill=blue!20,  opacity=0.5] ( 0.0,0.9) circle (1.0);
\draw[draw=darkgray] (-0.5,0.0) circle (1.0);
\draw[draw=darkgray] ( 0.5,0.0) circle (1.0);
\draw[draw=darkgray] ( 0.0,0.9) circle (1.0);

% Labels for sets
\node[align=center] at (-2.4,-0.6) {Llama3.1-8B\\[-0.2em](591)};
\node[align=center] at ( 2.4,-0.6) {GPT-oss-120B\\[-0.2em](1251)};
\node[align=center] at (-1.4, 1.6) {GPT-5.4\\[-0.2em](1496)};

% Region values
\node at (-0.9,-0.1) {5};
\node at ( 0.9,-0.1) {81};
\node at ( 0.0, 1.3) {\textbf{347}};
\node at (-0.6, 0.7) {4};
\node at ( 0.6, 0.7) {\textbf{588}};
\node at ( 0.0,-0.4) {25};
\node at ( 0.0, 0.3) {\textbf{557}};

\end{tikzpicture}
\caption{Venn diagram of applicable transformation rules.}
\label{fig:rules_overlap}
\Description{A three-set Venn diagram comparing the applicable
transformation rules identified by three language models:
Llama3.1-8B (591 rules), GPT-oss-120B (1251 rules), and GPT-5.4
(1496 rules). The center intersection contains 557 rules found by
all three models. The pairwise-only intersections contain 4
(Llama3.1-8B and GPT-5.4), 588 (GPT-oss-120B and GPT-5.4), and 25
(Llama3.1-8B and GPT-oss-120B). The model-specific regions contain
5, 81, and 347 rules, respectively.}
\end{figure}
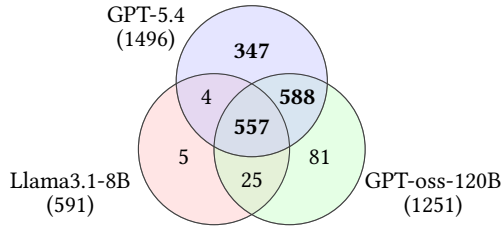
\camera{
\emph{\textbf{Semantic equivalence on Defects4J and BugsInPy.}} 
%To enhance the evaluation of the semantic correctness of transformations that did not produce an AST match, we replaced the corresponding source files with the generated transformations and executed the Defects4J and BugsInPy test suites. 
\autoref{tab:test-validation} shows that only a small fraction of these transformations resulted in successful repairs when there is no AST match. On average, GPT-oss-120B has the largest number of repair attempts on both Defects4J (200.5 attempted repairs per DSL) and BugsInPy (368.0), followed by GPT5.4 (127.8 and 328.8, respectively) and Llama3.1-8B (119.8 and 288.2, respectively). On Defects4J, the best outcome was achieved by GPT-oss-120B with GritQL, with 8 successful repairs (3.0\% success rate), while Ast-Grep attained the highest success rate of 6.7\% (6/89). On BugsInPy, substantially more successful repairs were observed, with Ast-Grep and GPT5.4 achieving the highest success rate of 17.9\% (38/212), and Comby (\#R) with Llama3.1-8B producing the most successful repairs (23/383). 
However, these results \reviewonly{are in a one-shot generation set up and} could be significanty improved with a feedback loop \cite{liu2025survey}. % implementing a  could possibly improve the Semantic equivalence.
}

\begin{tcolorbox}
\textbf{$\boldsymbol{RQ_3}$ insights and findings:}
GPT-5.4 achieves the best overall semantic correctness, particularly on complex transformations. GPT-oss-120B performs well in simpler settings, especially with Comby, but struggles with large edits. Llama3.1-8B shows consistently low performance, with many missing or incorrect transformations.
Qualitative analysis shows that GPT-5.4 generates more correct edits in complex scenarios, while smaller models often miss or incorrectly apply transformations.
\camera{Test-based validation further confirms that AST matching is a necessary condition for successful test passing, although only a subset of non AST-matching transformations repaired the bugs.}
Fully reproducing the \emph{intended} transformation remains the main challenge: even GPT-5.4 sometimes produces rules that are too specific to generalize or too partial to capture all required changes, and the gap between models increases on complex and non-local edits (e.g., multi-statement migrations).
\end{tcolorbox}

\subsection{RQ4: Complementarity}%: \red{{\small To what extent do different LLMs generate complementary transformation rules, and how much overlap exists between them?}}}

\autoref{fig:rules_overlap} presents the overlap of correctly inferred transformation rules across the three models in terms of AST match. Overall, GPT-based models substantially outperform Llama3.1-8B, which contributes very few unique correct transformations (5 cases). In contrast, GPT-5.4 identifies the largest number of unique rules (347), followed by GPT-oss-120B (81), highlighting their stronger capability to generalize beyond shared cases.

A large portion of rules is shared between GPT-5.4 and GPT-oss-120B (557+588), indicating that both models capture similar transformation patterns with different efficiency and verbosity, as discussed previously.
Our qualitative analysis shows that the set of rules on which all three models overlap corresponds to typically low-complexity cases, often reducible to one-line transformations.

Importantly, GPT-5.4 demonstrates a clear advantage in handling more complex transformations, as evidenced by its substantial number of unique solutions and its dominance in pairwise overlaps. This suggests that GPT-5.4 is better able to infer nuanced or structurally complex rules that are not captured by smaller models. In contrast, Llama3.1-8B is limited to simple patterns, such as one-line transformations, but remains a viable option for local deployment.

\begin{tcolorbox}
\textbf{$\boldsymbol{RQ_4}$ insights and findings:}
%\red{GPT-based models significantly outperform Llama3.1-8B in generating correct transformation rules. They not only achieve higher correctness, but also produce a substantially larger number of unique rules, demonstrating a stronger ability to generalize beyond shared patterns.
%
%Llama3.1-8B contributes very few unique transformations and is mostly limited to simple, one-line changes.}
GPT-5.4 is in a category of its own, with 347 unique correct rules that no other model produces, compared to 81 for GPT-oss-120B and only 5 for Llama3.1-8B.
 Our qualitative analysis confirms that only GPT-5.4 consistently solves complex transformations, while Llama3.1-8B is more limited but very effective for simple, one-line changes. % but remains a viable option for local deployment.
\end{tcolorbox}

\subsection{Threats to Validity}
\textbf{Prompt and LLM sensitivity.} Minor variations in prompt phrasing can significantly affect model outputs, influencing generated rule structure and DSL behavior, reducing reproducibility. To mitigate this, we standardize prompt templates across all experiments. 
Additionally, the temperature parameter can impact output variability and correctness. We mitigate this by fixing it to zero. 
Finally, the number and selection of pattern examples, particularly when using RAG for Comby, may bias a model toward specific rewriting strategies. To mitigate this, we use a fixed number of examples and consistent retrieval strategies across experiments.
These effects are compounded by differences in DSL expressiveness: some DSLs lack features such as guards or fine-grained node constraints, while others, like Ast-Grep, offer more advanced mechanisms for precisely targeting AST nodes, making direct comparisons challenging. To mitigate this, we evaluate three DSLs and constrain the LLM to use only the core match and rewrite functionalities of each DSL.

\invisibleskip % à fine tuner dans main.tex
\textbf{Evaluation metric limitations.}
The chosen evaluation metrics (e.g., rule application rate or syntactic correctness) may not fully capture semantic correctness. For instance, a transformation might fail syntactic checks (exact match or AST match) while still preserving semantic behavior when evaluated against a test suite. \camera{We investigated this limitation on the two benchmarks with available test suites, but the absence of systematically available executable oracles across all considered benchmarks prevents extending this evaluation more broadly, which is beyond the scope of this work.}

\invisibleskip % à fine tuner dans main.tex
\textbf{Scalability and complexity limits.}
The evaluation may suffer from bias on small or localized transformations. Performance could degrade on larger codebase, deeply nested structures, or transformations requiring global context. To mitigate this, we include datasets containing a diverse range of transformation complexities
%(e.g., Defects4J, BugsInPy, and PyMigBench),
which encompass both local and non-local changes.

\removed{\invisibleskip % à fine tuner dans main.tex
\textbf{Programming language bias.} The evaluation focuses on Python and Java, which do not cover the full spectrum of programming languages. However, these languages were selected to represent two distinct syntactic paradigms: indentation-sensitive languages (Python) and bracket-based languages (Java), which remain two of the most popular widespread languages.}

\invisibleskip % à fine tuner dans main.tex
\textbf{Programming language bias.} The evaluation is limited to Java and Python, two very popular languages representing two distinct syntactic paradigms (bracket-based and indentation-sensitive), but not the full spectrum of programming languages.

\invisibleskip % à fine tuner dans main.tex
\textbf{Dataset leakage.} There is a risk that models have been exposed to similar patterns or rules during training, potentially inflating performance. However, this threat is difficult to fully eliminate, as training data of proprietary LLMs is not publicly accessible.

\section{Recommendations}

Our results suggest recommendations for (i) practitioners choosing models and workflows for rule generation, (ii) researchers studying the scope and limits of rule synthesis, and (iii) transformation-language designers building more LLM-compatible tooling.

\structskip % à fine tuner dans main.tex
%\invisibleskip % à fine tuner dans main.tex
\textbf{For Practitioners.}
Transformation rule generation can help automate repetitive code modifications from minimal input. Developers can generate a rule from a single example and apply it recursively across a codebase, significantly reducing manual effort. Such techniques could be embedded into IDEs, where developers provide a representative code diff and receive suggested transformation rules that can be interactively reviewed and applied. An alternative usage scenario consists in generating transformation rules from an input code fragment and a natural-language instruction, enabling developers to describe the intended modification while leveraging LLMs to synthesize the corresponding rule.

However, our findings indicate that transformation rule generation is best suited for integration into \emph{semi-automated workflows} rather than fully autonomous refactoring systems. In particular, complex refactorings are better handled when decomposed into smaller, composable subrules, which improves both reliability and interpretability of the generated transformations. More generally, the intended level of abstraction is not always recoverable from examples alone: LLMs may overfit the observed edits, while more general and reusable rules often depend on explicit intent provided by the user or by additional specifications.

Finally, our results reveal an inherent trade-off between model capability and operational cost. Frontier models handle complex transformations better but require external APIs and higher computational resources. In contrast, smaller models such as Llama-based models are effective at generating simple, one-line transformation rules, making them suitable for local deployment scenarios where privacy constraints are critical. These results support hybrid workflows, combining different models depending on task complexity.

\structskip % à fine tuner dans main.tex
%\invisibleskip % à fine tuner dans main.tex
\textbf{For Researchers.}
Beyond benchmarking model performance, our results point to three important research directions: \emph{understanding which software-evolution tasks are inherently amenable to rule synthesis}, improving controllable generalization, and designing effective decomposition strategies for complex transformations.

In addition, rule generation can serve as a research instrument. Automatically inferred rules provide a powerful abstraction for analyzing software evolution at scale, enabling systematic studies of recurring transformation patterns. In particular, researchers can leverage these rules to analyze the distribution and frequency of common bug patterns, API misuses, and migration practices across projects.
Furthermore, transformation rules offer a structured and interpretable representation of LLM outputs, which can facilitate their analysis and evaluation. They can also be integrated into fuzzing pipelines, where generated rules are used to systematically explore variations of code transformations. Finally, rule mining can support dataset augmentation and the mining of software repositories, enabling richer and more diverse benchmarks.

Our results also raise questions about benchmark design. Datasets such as Defects4J, BugsInPy, and PyMigBench are significantly more challenging due to the size and complexity of their diffs. These transformations involve large, heterogeneous modifications that do not consistently share common structures between the ``before'' and ``after'' code, limiting the ability of models to infer concise and reusable transformation rules. These findings suggest that current benchmarks are not always aligned with the assumptions underlying transformation rule generation, and that future evaluations should explicitly account for task suitability.

\structskip % à fine tuner dans main.tex
%\invisibleskip % à fine tuner dans main.tex
\textbf{For Transformation-Language Designers and Tool Builders.}
We identify recurring failure modes that are relevant to DSL and tool design. Models may overgeneralize transformation rules, leading to excessive and potentially incorrect matches, or undergeneralize them, resulting in overly specific rules that fail to transfer to new contexts. This highlights the difficulty of balancing precision and generality when inferring transformation rules, and suggests the need for mechanisms to \emph{better control rule scope}. Transformation languages such as GritQL and Ast-Grep rely on AST node matching, which makes large rigid rules difficult to apply effectively; tool designers could consider supporting explicit rule decomposition or hierarchical rule composition. More generally, many remaining failures stem from syntactic and configuration constraints of the target DSL (e.g., malformed YAML\reviewonly{/TOML}, \reviewonly{missing rule separators,} invalid multi-node patterns) rather than from an inability to infer the intended edit. This suggests that more LLM-friendly configuration formats or better error recovery in DSL engines could substantially improve rule applicability.

Our results suggest that adaptation strategies for transformation-rule generation must be carefully aligned with the target DSL and software-evolution task. In our setting, RAG negatively impacted performance, likely because retrieved examples were biased toward Python transformations and unevenly relevant to the target context. More generally, external guidance only helps when it is relevant, diverse, and closely aligned with the requirements of the target transformation. These findings indicate that improving rule generation is not simply a matter of adding more context, but of better matching the conditioning signal to DSL constraints, task structure, and language diversity. We believe the qualitative analysis to be particularly useful in this regard, because it exposes recurring success patterns and failure anti-patterns that can be operationalized into more systematic support for future rule-generation systems.

\section{Related Work}
Recent research has explored automated generation of program transformation rules, either by learning from code edit examples or by inferring rules from recurring security and optimization patterns. In particular, many papers focus on generating rules from input and output code pairs resulting from one or more code changes. 

Ketkar et al.~\cite{ketkar_inferring_2022} infer transformation rules from type-change code edits using the Comby DSL and an anti-unification algorithm~\cite{plotkin_note_nodate}. This algorithm generalizes multiple code fragments by extracting their common structure and replacing differences with variable placeholders (e.g., Comby placeholders such as \texttt{:[var]}). This approach was later extended by PyEvolve~\cite{dilhara_pyevolve_2023}, which improves the handling of unseen data-flow and control-flow variants through graph-based analysis. More recently, PyCraft~\cite{dilhara_unprecedented_2024} further expands this work by introducing a Code Change Pattern (CPAT) miner to retrieve similar code changes for rule inference, combined with an LLM-based generator that produces synthetic CPAT variants.

Ramos et al. have developed MELT~\cite{ramos_melt_2023}, a rule generation tool that infers rules from pull requests with a relevant code edit, such as the API method we want to infer the rule from. They also proposed SPELL~\cite{ramos_spell_2026}, %which follows a similar approach to MELT but 
which augments MELT with synthetic input–output–context triplets generated using large language models. These triplets are fed to an anti-unification algorithm to infer an initial rule, which is then iteratively refined by an LLM-based agent.

However, SPELL %fundamentally 
differs from our setting in both data source and objective. First, SPELL relies on synthetically generated examples to construct transformation rules in a unique DSL, whereas our evaluation is conducted on real-world code changes. Second, SPELL assumes access to multiple examples of the same transformation, which are aggregated to infer a generalized rule that can later be applied to real-world repositories. 
In contrast we aim to infer transformation rules directly from real-world diffs across multiple DSLs.  %In contrast, our approach targets a more realistic scenario: we aim to infer transformation rules directly from real-world diffs, and apply them to other codebase.}
Because SPELL and MELT methods rely mainly on anti-unification and generation from simple examples to infer transformation rules, %%they are  %inherently 
%limited to code changes that share a common structural pattern. Consequently, 
they cannot capture transformations that involve some advanced semantic changes. % or lack sufficient syntactic similarity.
%For example, a code edit can correspond to a deletion, modification, or addition. However, 
Specifically, existing tools handle only deletions and modifications, as they rely on matching code fragments extracted from diffs. In contrast, our study evaluates LLMs on all types of code edits, including \emph{additions}. This increases the models' task complexity, since they must identify the appropriate anchor location in the code where the new statement should be inserted.

Besides, the automatic generation of rules is widely explored in fields such as the detection of recurrent security vulnerabilities and optimization issues. These approaches aim to identify common patterns in code and derive reusable transformations or fixes that can be systematically applied across a codebase. For instance, prior work like CQLLM~\cite{han_cqllm_2025} has leveraged LLMs and vector knowledge retrieval techniques to infer CodeQL transformation rules to detect common code vulnerabilities. Zhao et al.~\cite{zhao_semopt_2025} explored Semgrep rule generation from code diffs to generate optimization strategies which could be reused in a similar context at function-level. Wang et al. proposed RulePilot~\cite{wang_rulepilot_2025}, a solution using natural-language-based descriptions of a vulnerability to automatically generate the detection rules. They equipped the LLM with an abstract representation of the complexity of config rules into a standardized format, reducing hallucinations, allowing LLMs to focus on rule generation.

\camera{Nevertheless, these approaches are often limited for low-resource and domain-specific programming languages, which are underrepresented in current LLM training data~\cite{joel_survey_2025}. Researchers address these limitations with techniques such as domain-specific pre-training, fine-tuning, or retrieval-augmented generation (RAG).}

\section{Conclusion}
%In this paper, 
We presented an empirical study on the ability of LLMs to generate code transformation rules expressed in DSLs. By evaluating three LLMs across three transformation DSLs and six benchmarks covering multiple software-evolution tasks, we provide a comprehensive assessment of rule synthesis capabilities in realistic settings.

Our results show that LLMs are capable of generating syntactically valid transformation rules, with strong performance though not perfect, from models such as GPT-5.4. However, effectiveness varies significantly depending on the complexity of the transformation, the structure of the underlying code changes, and the expressiveness of the target DSL. In particular, localized edits such as API misuse corrections are substantially more amenable to rule synthesis than function-level repairs or multi-statement migrations. We also observe a trade-off between abstraction and coverage, where models tend to generate overly specific rules instead of leveraging meta-variables and higher-level abstractions. %, though explicitness sometimes better captures complex transformations. 
\camera{At the same time, many of the generated rules exhibit high reusability, successfully matching multiple code instances across the evaluated datasets.} Moreover, our qualitative analysis explains the concrete mechanisms behind model differences and failure modes, including formatting failures, over- and under-generalization, and missing edits. \camera{The anti-unification baseline performs overall worse than LLMs in terms of both syntactic and semantic correctness. Nevertheless, combining anti-unification with frontier LLMs could provide a promising hybrid approach with improved determinism.}

Overall, our findings show that the main question has shifted from whether LLMs can synthesize transformation rules to under which task and representation conditions the generated rules are sound, correct, generalizable, and reusable. Such approaches are better suited for integration into semi-automated workflows, where generated rules can be reviewed, refined, and applied at scale.

%Our full experimental pipeline is publicly available and can be executed within a relatively short time frame (approximately 2--3 days per model), making it a reusable contribution for the community to reproduce our results, 

We plan to extend the evaluation to additional benchmarks, models, and DSLs, and investigate possible improvements. 
A natural next step is to leverage this capability in real-world workflows, e.g., by integrating rule generation into IDEs or CI pipelines. Beyond deployment, future work also includes designing refinement loops with iterative feedback to improve incorrect rules (if any).

\structskip
\textbf{Data Availability Statement.}
We provide both a result artifact on Figshare~\cite{figshare-artifact} and a replication package
\url{https://anonymous.4open.science/r/EmpiricalStudyTransformationRules-13B3}.
\begin{acks}
This work is supported by the Inria Défi LLM4Code (DGDS012482).
\end{acks}

\bibliographystyle{ACM-Reference-Format}
\bibliography{references}

@inproceedings{islam_pymigbench_2023,
	address = {Melbourne, Australia},
	title = {{PyMigBench}: {A} {Benchmark} for {Python} {Library} {Migration}},
	copyright = {https://doi.org/10.15223/policy-029},
	isbn = {979-8-3503-1184-6},
	shorttitle = {{PyMigBench}},
	url = {https://ieeexplore.ieee.org/document/10174111/},
	doi = {10.1109/MSR59073.2023.00075},
	language = {en},
	urldate = {2025-11-05},
	booktitle = {2023 {IEEE}/{ACM} 20th {International} {Conference} on {Mining} {Software} {Repositories} ({MSR})},
	publisher = {IEEE},
	author = {Islam, Mohayeminul and Jha, Ajay Kumar and Nadi, Sarah and Akhmetov, Ildar},
	month = may,
	year = {2023},
	pages = {511--515},
}

@misc{zhao_semopt_2025,
	title = {{SemOpt}: {LLM}-{Driven} {Code} {Optimization} via {Rule}-{Based} {Analysis}},
	shorttitle = {{SemOpt}},
	url = {http://arxiv.org/abs/2510.16384},
	doi = {10.48550/arXiv.2510.16384},
	language = {en},
	urldate = {2025-11-20},
	publisher = {arXiv},
	author = {Zhao, Yuwei and Xiao, Yuan-An and Xiao, Qianyu and Zhang, Zhao and Xiong, Yingfei},
	month = oct,
	year = {2025},
	note = {arXiv:2510.16384 [cs]},
}

@inproceedings{ramos_melt_2023,
	title = {{MELT}: {Mining} {Effective} {Lightweight} {Transformations} from {Pull} {Requests}},
	issn = {2643-1572},
	shorttitle = {{MELT}},
	url = {https://ieeexplore.ieee.org/document/10298355/},
	doi = {10.1109/ASE56229.2023.00117},
	urldate = {2025-12-18},
	booktitle = {2023 38th {IEEE}/{ACM} {International} {Conference} on {Automated} {Software} {Engineering} ({ASE})},
	author = {Ramos, Daniel and Mitchell, Hailie and Lynce, Inês and Manquinho, Vasco and Martins, Ruben and Goues, Claire Le},
	month = sep,
	year = {2023},
	pages = {1516--1528},
}

@inproceedings{dilhara_pyevolve_2023,
	address = {Melbourne, Australia},
	title = {{PYEVOLVE}: {Automating} {Frequent} {Code} {Changes} in {Python} {ML} {Systems}},
	isbn = {978-1-6654-5701-9},
	shorttitle = {{PYEVOLVE}},
	url = {https://ieeexplore.ieee.org/document/10172702/},
	doi = {10.1109/ICSE48619.2023.00091},
	language = {en},
	urldate = {2025-12-19},
	booktitle = {2023 {IEEE}/{ACM} 45th {International} {Conference} on {Software} {Engineering} ({ICSE})},
	publisher = {IEEE},
	author = {Dilhara, Malinda and Dig, Danny and Ketkar, Ameya},
	month = may,
	year = {2023},
	pages = {995--1007},
}

@article{dilhara_unprecedented_2024,
	title = {Unprecedented {Code} {Change} {Automation}: {The} {Fusion} of {LLMs} and {Transformation} by {Example}},
	volume = {1},
	copyright = {https://creativecommons.org/licenses/by/4.0/},
	issn = {2994-970X},
	shorttitle = {Unprecedented {Code} {Change} {Automation}},
	url = {https://dl.acm.org/doi/10.1145/3643755},
	doi = {10.1145/3643755},
	language = {en},
	number = {FSE},
	urldate = {2025-12-19},
	journal = {Proceedings of the ACM on Software Engineering},
	author = {Dilhara, Malinda and Bellur, Abhiram and Bryksin, Timofey and Dig, Danny},
	month = jul,
	year = {2024},
	pages = {631--653},
}

@inproceedings{van_tonder_lightweight_2019,
	address = {New York, NY, USA},
	series = {{PLDI} 2019},
	title = {Lightweight multi-language syntax transformation with parser parser combinators},
	isbn = {978-1-4503-6712-7},
	url = {https://dl.acm.org/doi/10.1145/3314221.3314589},
	doi = {10.1145/3314221.3314589},
	urldate = {2025-12-22},
	booktitle = {Proceedings of the 40th {ACM} {SIGPLAN} {Conference} on {Programming} {Language} {Design} and {Implementation}},
	publisher = {Association for Computing Machinery},
	author = {van Tonder, Rijnard and Le Goues, Claire},
	month = jun,
	year = {2019},
	pages = {363--378},
}

@inproceedings{ketkar_inferring_2022,
	address = {New York, NY, USA},
	series = {{ICSE} '22},
	title = {Inferring and applying type changes},
	isbn = {978-1-4503-9221-1},
	url = {https://dl.acm.org/doi/10.1145/3510003.3510115},
	doi = {10.1145/3510003.3510115},
	urldate = {2025-12-22},
	booktitle = {Proceedings of the 44th {International} {Conference} on {Software} {Engineering}},
	publisher = {Association for Computing Machinery},
	author = {Ketkar, Ameya and Smirnov, Oleg and Tsantalis, Nikolaos and Dig, Danny and Bryksin, Timofey},
	month = jul,
	year = {2022},
	pages = {1206--1218},
}

@misc{wang_rulepilot_2025,
	title = {{RulePilot}: {An} {LLM}-{Powered} {Agent} for {Security} {Rule} {Generation}},
	shorttitle = {{RulePilot}},
	url = {http://arxiv.org/abs/2511.12224},
	doi = {10.48550/arXiv.2511.12224},
	language = {en},
	urldate = {2025-12-31},
	publisher = {arXiv},
	author = {Wang, Hongtai and Xu, Ming and Guo, Yanpei and Han, Weili and Lim, Hoon Wei and Dong, Jin Song},
	month = nov,
	year = {2025},
	note = {arXiv:2511.12224 [cs]},
}

@misc{han_cqllm_2025,
	title = {{CQLLM}: {A} {Framework} for {Generating} {CodeQL} {Security} {Vulnerability} {Detection} {Code} {Based} on {Large} {Language} {Model}},
	shorttitle = {{CQLLM}},
	url = {https://www.preprints.org/manuscript/202510.1458},
	doi = {10.20944/preprints202510.1458.v1},
	language = {en},
	urldate = {2025-12-31},
	publisher = {Preprints},
	author = {Han, Weihong and Chen, Chan and Zhu, Junyi and Zhan, Rufeng and Han, Weihong},
	month = oct,
	year = {2025},
}

@inproceedings{galappaththi_empirical_2024,
	address = {Barcelona Spain},
	title = {An {Empirical} {Study} of {API} {Misuses} of {Data}-{Centric} {Libraries}},
	isbn = {979-8-4007-1047-6},
	url = {https://dl.acm.org/doi/10.1145/3674805.3686685},
	doi = {10.1145/3674805.3686685},
	language = {en},
	urldate = {2026-01-19},
	booktitle = {Proceedings of the 18th {ACM}/{IEEE} {International} {Symposium} on {Empirical} {Software} {Engineering} and {Measurement}},
	publisher = {ACM},
	author = {Galappaththi, Akalanka and Nadi, Sarah and Treude, Christoph},
	month = oct,
	year = {2024},
	pages = {245--256},
}

@inproceedings{karampatsis_how_2020,
	address = {New York, NY, USA},
	series = {{MSR} '20},
	title = {How {Often} {Do} {Single}-{Statement} {Bugs} {Occur}? {The} {ManySStuBs4J} {Dataset}},
	isbn = {978-1-4503-7517-7},
	shorttitle = {How {Often} {Do} {Single}-{Statement} {Bugs} {Occur}?},
	url = {https://dl.acm.org/doi/10.1145/3379597.3387491},
	doi = {10.1145/3379597.3387491},
	urldate = {2026-01-22},
	booktitle = {Proceedings of the 17th {International} {Conference} on {Mining} {Software} {Repositories}},
	publisher = {Association for Computing Machinery},
	author = {Karampatsis, Rafael-Michael and Sutton, Charles},
	month = sep,
	year = {2020},
	pages = {573--577},
}

@inproceedings{widyasari_bugsinpy_2020,
	title = {{BugsInPy}: {A} {Database} of {Existing} {Bugs} in {Python} {Programs} to {Enable} {Controlled} {Testing} and {Debugging} {Studies}},
	shorttitle = {{BugsInPy}},
	url = {http://arxiv.org/abs/2401.15481},
	doi = {10.1145/3368089.3417943},
	language = {en},
	urldate = {2026-02-06},
	booktitle = {Proceedings of the 28th {ACM} {Joint} {Meeting} on {European} {Software} {Engineering} {Conference} and {Symposium} on the {Foundations} of {Software} {Engineering}},
	author = {Widyasari, Ratnadira and Sim, Sheng Qin and Lok, Camellia and Qi, Haodi and Phan, Jack and Tay, Qijin and Tan, Constance and Wee, Fiona and Tan, Jodie Ethelda and Yieh, Yuheng and Goh, Brian and Thung, Ferdian and Kang, Hong Jin and Hoang, Thong and Lo, David and Ouh, Eng Lieh},
	month = nov,
	year = {2020},
	note = {arXiv:2401.15481 [cs]},
	pages = {1556--1560},
}

@misc{ramos_spell_2026,
	title = {{SPELL}: {Synthesis} of {Programmatic} {Edits} using {LLMs}},
	shorttitle = {{SPELL}},
	url = {http://arxiv.org/abs/2602.01107},
	doi = {10.48550/arXiv.2602.01107},
	language = {en},
	urldate = {2026-02-09},
	publisher = {arXiv},
	author = {Ramos, Daniel and Gamboa, Catarina and Lynce, Inês and Manquinho, Vasco and Martins, Ruben and Goues, Claire Le},
	month = feb,
	year = {2026},
	note = {arXiv:2602.01107 [cs]},
}

@inproceedings{just_defects4j_2014,
	address = {New York, NY, USA},
	series = {{ISSTA} 2014},
	title = {{Defects4J}: a database of existing faults to enable controlled testing studies for {Java} programs},
	isbn = {978-1-4503-2645-2},
	shorttitle = {{Defects4J}},
	url = {https://dl.acm.org/doi/10.1145/2610384.2628055},
	doi = {10.1145/2610384.2628055},
	urldate = {2026-02-12},
	booktitle = {Proceedings of the 2014 {International} {Symposium} on {Software} {Testing} and {Analysis}},
	publisher = {Association for Computing Machinery},
	author = {Just, René and Jalali, Darioush and Ernst, Michael D.},
	month = jul,
	year = {2014},
	pages = {437--440},
}

@misc{amin_jmigbench_2026,
	title = {{JMigBench}: {A} {Benchmark} for {Evaluating} {LLMs} on {Source} {Code} {Migration} ({Java} 8 to {Java} 11)},
	shorttitle = {{JMigBench}},
	url = {http://arxiv.org/abs/2602.09930},
	doi = {10.48550/arXiv.2602.09930},
	urldate = {2026-02-18},
	publisher = {arXiv},
	author = {Amin, Nishil and Fei, Zhiwei and Li, Xiang and Petke, Justyna and Ye, He},
	month = feb,
	year = {2026},
	note = {arXiv:2602.09930 [cs]},
}

@misc{joel_survey_2025,
	title = {A {Survey} on {LLM}-based {Code} {Generation} for {Low}-{Resource} and {Domain}-{Specific} {Programming} {Languages}},
	url = {http://arxiv.org/abs/2410.03981},
	doi = {10.48550/arXiv.2410.03981},
	language = {en},
	urldate = {2026-06-29},
	publisher = {arXiv},
	author = {Joel, Sathvik and Wu, Jie JW and Fard, Fatemeh H.},
	month = sep,
	year = {2025},
	note = {arXiv:2410.03981 [cs.SE]},
}

@misc{baltes_guidelines_2026,
  title = {Guidelines for Empirical Studies in Software Engineering involving Large Language Models},
  url = {https://arxiv.org/abs/2508.15503},
  doi = {10.48550/arXiv.2508.15503},
  urldate = {2026-07-15},
  publisher = {arXiv},
  author = {Baltes, Sebastian and Angermeir, Florian and Arora, Chetan and Muñoz Barón, Marvin and Chen, Chunyang and Böhme, Lukas and Calefato, Fabio and Ernst, Neil and Falessi, Davide and Fitzgerald, Brian and Fucci, Davide and He, Junda and Treude, Christoph and Kalinowski, Marcos and Lambiase, Stefano and Russo, Daniel and Lungu, Mircea and Martinez Montes, Cristina and Prechelt, Lutz and Ralph, Paul and van Tonder, Rijnard and Wagner, Stefan},
  month = jun,
  year = {2026},
  note = {Accepted manuscript, arXiv:2508.15503v7 [cs.SE]}
}

@misc{astgrep2026,
  author       = {{ast-grep}},
  title        = {ast-grep},
  year         = {2026},
  url          = {https://ast-grep.github.io/},
  lastaccessed = {July 30, 2026}
}

@misc{gritql2026,
  author       = {{GritQL}},
  title        = {GritQL},
  year         = {2026},
  url          = {https://docs.grit.io/},
  lastaccessed = {July 30, 2026}
}

@article{plotkin_note_nodate,
	title = {A {Note} on {Inductive} {Generalization}},
	language = {en},
    year = {1970},
    journal = {...},
	author = {Plotkin, Gordon D},
}

@dataset{figshare-artifact,
  author = {Axel Allain},
  title = {Result Artifact for ``Code Transformation Rule Synthesis Using LLMs: Potential and Limits''},
  year = {2026},
  publisher = {Figshare},
  version = {1},
  doi = {10.6084/m9.figshare.31861126.v1},
  url = {https://doi.org/10.6084/m9.figshare.31861126.v1}
}

@inproceedings{siy1998challenges,
  title={Challenges in evolving a large scale software product},
  author={Siy, Harvey P and Perry, Dewayne E},
  booktitle={Proceedings of Principles of Software Evolution Workshop at the International Software Engineering Conference, ICSE},
  volume={98},
  pages={251--260},
  year={1998}
}

@inproceedings{sarkar2009software,
  title={Software challenges in extreme scale systems},
  author={Sarkar, Vivek and Harrod, William and Snavely, Allan E},
  booktitle={Journal of Physics: Conference Series},
  volume={180},
  number={1},
  pages={012045},
  year={2009},
	doi = {10.1088/1742-6596/180/1/012045},
}

@article{northrop2006ultra,
  title={Ultra-large-scale systems: The software challenge of the future},
  author={Northrop, Linda and Feiler, Peter and Gabriel, Richard P and Goodenough, John and Linger, Rick and Longstaff, Tom and Kazman, Rick and Klein, Mark and Sullivan, Kevin and Wallnau, Kurt and others},
  year={2006}
}

@article{glass2001frequently,
	title        = {Frequently forgotten fundamental facts about software engineering},
	author       = {Glass, Robert L},
	year         = 2001,
	journal      = {IEEE software},
	volume       = 18,
	number       = 3,
	pages        = {112--111},
	doi = {10.1109/MS.2001.922739},
}

@article{alkhatib1992maintenance,
	title        = {The maintenance problem of application software: An empirical analysis},
	author       = {Alkhatib, Ghazi},
	year         = 1992,
	journal      = {Journal of Software Maintenance: Research and Practice},
	publisher    = {Wiley Online Library},
	volume       = 4,
	number       = 2,
	pages        = {83--104},
    doi = {10.1002/smr.4360040203}
}

@article{banker1993software,
  author  = {Banker, Rajiv D. and Datar, Srikant M. and Kemerer, Chris F. and Zweig, Dani},
  title   = {Software Complexity and Maintenance Costs},
  journal = {Communications of the ACM},
  volume  = {36},
  number  = {11},
  pages   = {81--94},
  year    = {1993},
  doi     = {10.1145/163359.163375}
}

@article{le2019automated,
  title={Automated program repair},
  author={Le Goues, Claire and Pradel, Michael and Roychoudhury, Abhik},
  journal={Communications of the ACM},
  volume={62},
  number={12},
  pages={56--65},
  year={2019},
  publisher={ACM New York, NY, USA},
	doi = {10.1145/3318162},
}

@article{liu2021critical,
  title={A critical review on the evaluation of automated program repair systems},
  author={Liu, Kui and Li, Li and Koyuncu, Anil and Kim, Dongsun and Liu, Zhe and Klein, Jacques and Bissyand{\'e}, Tegawend{\'e} F},
  journal={Journal of Systems and Software},
  volume={171},
  pages={110817},
  year={2021},
  publisher={Elsevier},
	doi = {10.1016/j.jss.2020.110817},
}

@phdthesis{monperrus2018living,
  title={The living review on automated program repair},
  author={Monperrus, Martin},
  year={2018},
  school={HAL Archives Ouvertes}
}

@inproceedings{golubev2021one,
  title={One thousand and one stories: a large-scale survey of software refactoring},
  author={Golubev, Yaroslav and Kurbatova, Zarina and AlOmar, Eman Abdullah and Bryksin, Timofey and Mkaouer, Mohamed Wiem},
  booktitle={Proceedings of the 29th ACM joint meeting on european software engineering conference and symposium on the foundations of software engineering},
  pages={1303--1313},
  year={2021},
	doi = {10.1145/3468264.3473924},
}

@article{lacerda2020code,
  title={Code smells and refactoring: A tertiary systematic review of challenges and observations},
  author={Lacerda, Guilherme and Petrillo, Fabio and Pimenta, Marcelo and Gu{\'e}h{\'e}neuc, Yann Ga{\"e}l},
  journal={Journal of Systems and Software},
  volume={167},
  pages={110610},
  year={2020},
  publisher={Elsevier},
	doi = {10.1016/j.jss.2020.110610},
}

@article{mens2004survey,
  title={A survey of software refactoring},
  author={Mens, Tom and Tourw{\'e}, Tom},
  journal={IEEE Transactions on software engineering},
  volume={30},
  number={2},
  pages={126--139},
  year={2004},
  publisher={IEEE},
	doi = {10.1109/TSE.2004.1265817},
}

@inproceedings{amann2016mubench,
  title={MUBench: A benchmark for API-misuse detectors},
  author={Amann, Sven and Nadi, Sarah and Nguyen, Hoan A and Nguyen, Tien N and Mezini, Mira},
  booktitle={Proceedings of the 13th international conference on mining software repositories},
  pages={464--467},
  year={2016},
  doi={10.1145/2901739.2903506}
}

@article{amann2018systematic,
  title={A systematic evaluation of static api-misuse detectors},
  author={Amann, Sven and Nguyen, Hoan Anh and Nadi, Sarah and Nguyen, Tien N and Mezini, Mira},
  journal={IEEE Transactions on Software Engineering},
  volume={45},
  number={12},
  pages={1170--1188},
  year={2018},
  publisher={IEEE},
  doi = {10.48550/arXiv.1712.00242}
}

@inproceedings{sven2019investigating,
  title={Investigating next steps in static API-misuse detection},
  author={Sven, Amann and Nguyen, Hoan Anh and Nadi, Sarah and Nguyen, Tien N and Mezini, Mira},
  booktitle={2019 IEEE/ACM 16th International Conference on Mining Software Repositories (MSR)},
  pages={265--275},
  year={2019},
  organization={IEEE},
	doi = {10.1109/MSR.2019.00053},
}

@inproceedings{li2021large,
  title={A Large-scale Study on API Misuses in the Wild},
  author={Li, Xia and Jiang, Jiajun and Benton, Samuel and Xiong, Yingfei and Zhang, Lingming},
  booktitle={2021 14th IEEE conference on software testing, verification and validation (ICST)},
  pages={241--252},
  year={2021},
  organization={IEEE},
	doi = {10.1109/ICST49551.2021.00034},
}

@inproceedings{nguyen2016mapping,
  title={Mapping API elements for code migration with vector representations},
  author={Nguyen, Trong Duc and Nguyen, Anh Tuan and Nguyen, Tien N},
  booktitle={Proceedings of the 38th international conference on software engineering companion},
  pages={756--758},
  year={2016},
	doi = {10.1145/2889160.2892661},
}

@inproceedings{khelladi2020co,
  title={Co-evolving code with evolving metamodels},
  author={Khelladi, Djamel Eddine and Combemale, Benoit and Acher, Mathieu and Barais, Olivier and J{\'e}z{\'e}quel, Jean-Marc},
  booktitle={Proceedings of the ACM/IEEE 42nd international conference on software engineering},
  pages={1496--1508},
  year={2020},
	doi = {10.1145/3377811.3380324},
}

@inproceedings{le2021untangling,
  title={Untangling spaghetti of evolutions in software histories to identify code and test co-evolutions},
  author={Le Dilavrec, Quentin and Khelladi, Djamel Eddine and Blouin, Arnaud and J{\'e}z{\'e}quel, Jean-Marc},
  booktitle={2021 IEEE International Conference on Software Maintenance and Evolution (ICSME)},
  pages={206--216},
  year={2021},
  organization={IEEE},
	doi = {10.1109/ICSME52107.2021.00025},
}

@article{kebaili2025automated,
  title={Automated co-evolution of metamodels and code},
  author={Kebaili, Zohra Kaouter and Khelladi, Djamel Eddine and Acher, Mathieu and Barais, Olivier},
  journal={IEEE Transactions on Software Engineering},
  year={2025},
  publisher={IEEE},
	doi = {10.1109/TSE.2025.3540545},
}

@article{miranda2025test,
  title={Test Co-Evolution in Software Projects: A Large-Scale Empirical Study},
  author={Miranda, Charles and Avelino, Guilherme and Santos Neto, Pedro},
  journal={Journal of Software: Evolution and Process},
  volume={37},
  number={7},
  pages={e70035},
  year={2025},
  publisher={Wiley Online Library},
	doi = {10.1002/smr.70035},
}

@article{cordeiro2024empirical,
  title={An empirical study on the code refactoring capability of large language models},
  author={Cordeiro, Jonathan and Noei, Shayan and Zou, Ying},
  journal={ACM Transactions on Software Engineering and Methodology},
  year={2024},
  publisher={ACM New York, NY},
  doi={10.1145/3801158}
}

@inproceedings{ziftci2025migrating,
  title={Migrating code at scale with llms at google},
  author={Ziftci, Celal and Nikolov, Stoyan and Sj{\"o}vall, Anna and Kim, Bo and Codecasa, Daniele and Kim, Max},
  booktitle={Proceedings of the 33rd ACM International Conference on the Foundations of Software Engineering},
  pages={162--173},
  year={2025},
	doi = {10.48550/arXiv.2504.09691},
}

@inproceedings{zine2025llm,
  title={LLM-based Co-Evolution of Configurable Software Systems},
  author={Zine, Nada and Quinton, Cl{\'e}ment and Rouvoy, Romain},
  booktitle={Proceedings of the 2025 29th ACM International Systems and Software Product Line Conference-Volume A},
  pages={27--38},
  year={2025},
	doi = {10.1145/3744915.3748460},
}

@inproceedings{zhang2023multilingual,
  author    = {Zhang, Jiyang and Nie, Pengyu and Li, Junyi Jessy and Gligoric, Milos},
  title     = {Multilingual Code Co-Evolution Using Large Language Models},
  booktitle = {Proceedings of the 31st ACM Joint European Software Engineering Conference and Symposium on the Foundations of Software Engineering},
  pages     = {695--707},
  year      = {2023},
  doi       = {10.1145/3611643.3616350}
}

@inproceedings{bouzenia2025repairagent,
  title={Repairagent: An autonomous, llm-based agent for program repair},
  author={Bouzenia, Islem and Devanbu, Premkumar and Pradel, Michael},
  booktitle={2025 IEEE/ACM 47th International Conference on Software Engineering (ICSE)},
  pages={2188--2200},
  year={2025},
  organization={IEEE},
  doi={10.1109/ICSE55347.2025.00157}
}

@inproceedings{jin2023inferfix,
  title={Inferfix: End-to-end program repair with llms},
  author={Jin, Matthew and Shahriar, Syed and Tufano, Michele and Shi, Xin and Lu, Shuai and Sundaresan, Neel and Svyatkovskiy, Alexey},
  booktitle={Proceedings of the 31st ACM joint european software engineering conference and symposium on the foundations of software engineering},
  pages={1646--1656},
  year={2023},
  doi = {10.1145/3611643.3613892},
}

@article{yang2025survey,
  title={A survey of LLM-based automated program repair: Taxonomies, design paradigms, and applications},
  author={Yang, Boyang and Cai, Zijian and Liu, Fengling and Le, Bach and Zhang, Lingming and Bissyand{\'e}, Tegawend{\'e} F and Liu, Yang and Tian, Haoye},
  journal={arXiv preprint arXiv:2506.23749},
  year={2025},
	doi = {10.48550/arXiv.2506.23749},
}

@inproceedings{cordeiro2025llm,
  title={LLM-Driven Code Refactoring: Opportunities and Limitations},
  author={Cordeiro, Jonathan and Noei, Shayan and Zou, Ying},
  booktitle={2025 IEEE/ACM Second IDE Workshop (IDE)},
  pages={32--36},
  year={2025},
  organization={IEEE},
}

@inproceedings{shirafuji2023refactoring,
  title={Refactoring programs using large language models with few-shot examples},
  author={Shirafuji, Atsushi and Oda, Yusuke and Suzuki, Jun and Morishita, Makoto and Watanobe, Yutaka},
  booktitle={2023 30th Asia-Pacific Software Engineering Conference (APSEC)},
  pages={151--160},
  year={2023},
  organization={IEEE},
	doi = {10.1109/APSEC60848.2023.00025},
}

@inproceedings{liu2025survey,
  title={A survey on the feedback mechanism of LLM-based AI agents},
  author={Liu, Zhipeng and Bai, Xuefeng and Chen, Kehai and Chen, Xinyang and Li, Xiucheng and Xiang, Yang and Liu, Jin and Li, Hong-Dong and Wang, Yaowei and Nie, Liqiang and others},
  booktitle={Proceedings of the Thirty-Fourth International Joint Conference on Artificial Intelligence},
  pages={10582--10592},
  year={2025},
  organization={International Joint Conferences on Artificial Intelligence},
  doi={10.24963/ijcai.2025/1175}
}

\end{document}